\documentclass[12pt,onecolumn]{IEEEtran}
\usepackage{makecell,amsbsy,amsmath,amssymb,epsfig,bbm,mathrsfs,multirow,amsthm,bm}
\usepackage{array,multirow,graphicx}
\usepackage[english]{babel}
\usepackage[justification=centering]{caption}
\usepackage{url}
\usepackage{threeparttable}
\usepackage{epstopdf}
\usepackage{subfigure}
\usepackage[ruled,linesnumbered]{algorithm2e}
\usepackage{color,xcolor}
\usepackage{hyperref}

\DeclareMathAlphabet{\mathpzc}{OT1}{pzc}{m}{it}

\usepackage{url}

\IEEEoverridecommandlockouts

\begin{document}
\bibliographystyle{IEEE2}

\title{Securing Large-Scale D2D Networks Using Covert Communication and Friendly Jamming}

\author{Shaohan Feng,~\IEEEmembership{Member,~IEEE}, Xiao Lu,~\IEEEmembership{Member,~IEEE}, \\Sumei Sun,~\IEEEmembership{Fellow,~IEEE}, Dusit Niyato,~\IEEEmembership{Fellow,~IEEE}, \\ and Ekram Hossain,~\IEEEmembership{Fellow,~IEEE}\thanks{S. Feng and S. Sun are with the Institute for Infocomm Research, Singapore. Email: \{fengs, sunsm\}@i2r.a-star.edu.sg.}
\thanks{X. Lu is with York University, Toronto, ON M3J 1P3, Canada. Email: luxiao@yorku.ca.} 
\thanks{D. Niyato is with the School of Computer Science and Engineering, Nanyang Technological University, Singapore. Email: DNIYATO@ntu.edu.sg.} 
\thanks{E. Hossain is with the Department of Electrical and Computer Engineering, University of Manitoba, Canada. Email: Ekram.Hossain@umanitoba.ca.}}

\maketitle

\vspace{-15mm}

\begin{abstract}\vspace{-2mm}
We exploit both covert communication and friendly jamming to propose a friendly jamming-assisted covert communication and use it to doubly secure a large-scale device-to-device (D2D) network against eavesdroppers (i.e., wardens). 
The D2D transmitters defend against the wardens by: 1) hiding their transmissions with enhanced covert communication, and 2) leveraging friendly jamming to ensure information secrecy even if the D2D transmissions are detected. We model the combat between the wardens and the D2D network (the transmitters and the friendly jammers) as a two-stage Stackelberg game. Therein, the wardens are the followers at the lower stage aiming to minimize their detection errors, and the D2D network is the leader at the upper stage aiming to maximize its utility (in terms of link reliability and communication security) subject to the constraint on communication covertness. We apply stochastic geometry to model the network spatial configuration so as to conduct a system-level study. We develop a bi-level optimization algorithm to search for the equilibrium of the proposed Stackelberg game based on the successive convex approximation (SCA) method and Rosenbrock method. Numerical results reveal interesting insights. We observe that without the assistance from the jammers, it is difficult to achieve covert communication on D2D transmission. Moreover, we illustrate the advantages of the proposed friendly jamming-assisted covert communication by comparing it with the information-theoretical secrecy approach in terms of the secure communication probability and network utility.
\end{abstract}



\section{Introduction}
\label{sec:introduction}

\subsection{Background and Motivation}

Recently, D2D communication has attracted great attention due to its capability of enabling direct communication among proximal mobile devices without the involvement of network infrastructure such as a base station (BS) and the core network~\cite{9169911}. By such, D2D communication has emerged as a promising solution to the provision of energy-efficient communication with low end-to-end latency in supporting miscellaneous location-based peer-to-peer applications~\cite{8474350}. However, the D2D communications are vulnerable to malicious attacks such as eavesdropping due to the broadcast nature of the wireless medium. In general, it is challenging to secure D2D communication due to the following reasons:
\begin{itemize}
\item In various application scenarios, the D2D devices are resource-constrained and hence cannot afford high overhead in computation and communication. In this case, heavy resource-dependent methods such as the traditional cryptographic methods may not be feasible~\cite{9169911}.

\item The  D2D networks may not have any centralized support to execute functionalities such as authentication to achieve confidentiality of the transmissions.

\item Merely achieving a positive rate difference between legitimate and wiretapping channels is not sufficient from the perspective of privacy protection. For example, in a military application, it may be required to prevent the enemy from detecting transmissions so as to hide the presence and activities of the legitimate army~\cite{7805182}.
\end{itemize}



\begin{figure}[!]
	\centering
	\includegraphics[width=1\textwidth,trim=80 160 250 70, clip]{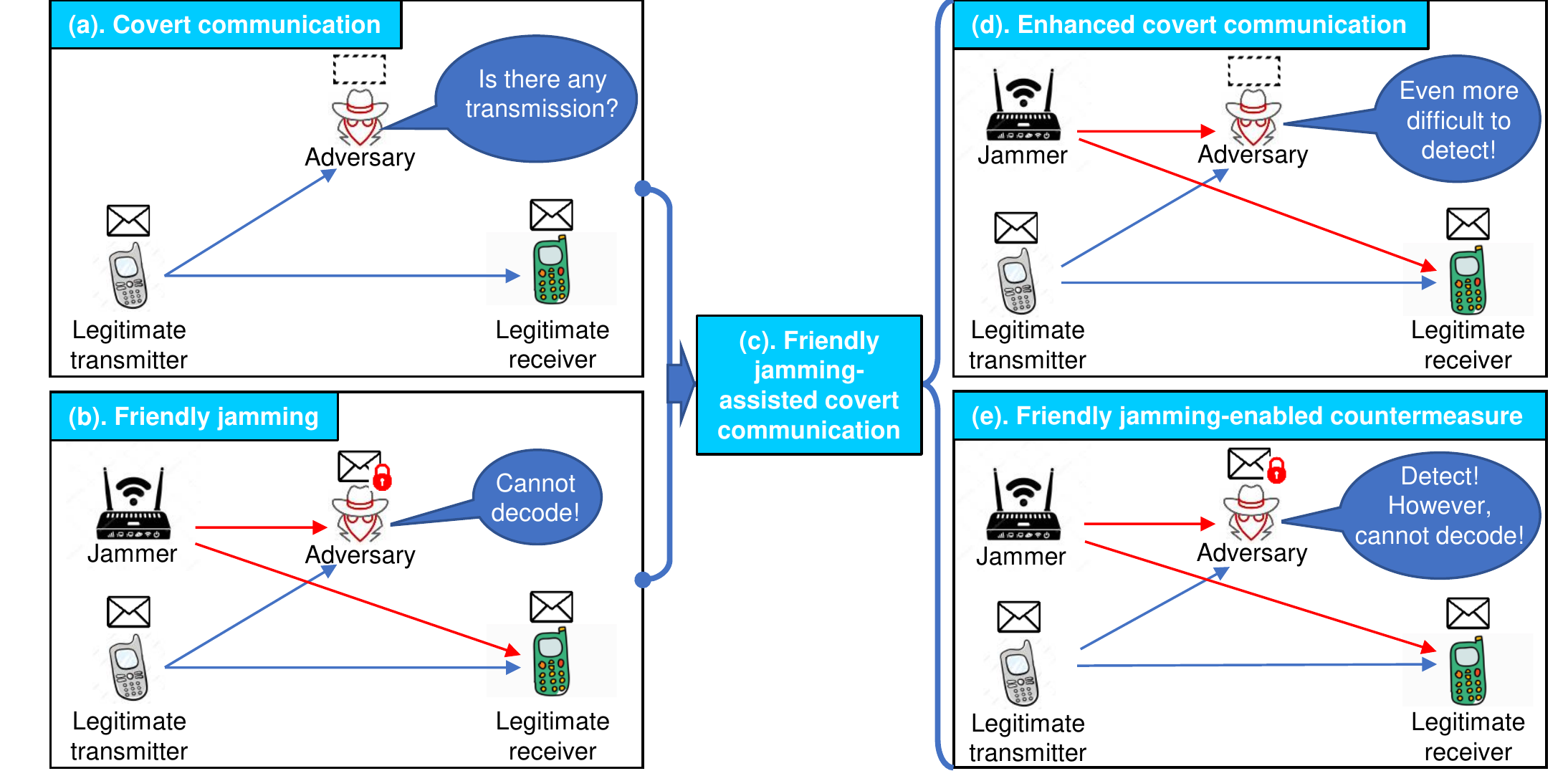}
	\caption{(a) Covert communication, (b) friendly jamming, and (c) friendly jamming-assisted covert communication including (d) enhanced covert communication and (e) friendly jamming-enabled countermeasure.}
	\label{fig:covert_enhanced_countermeasure}
\end{figure}

The above context raises the need for covert communication~\cite{9736993} (also referred to as undetectable communication), which is an enabler for reliable transmission from a legitimate transmitter to its intended receiver while undetectable by a vigilant adversary as shown in Fig.~\ref{fig:covert_enhanced_countermeasure}(a). Covert communication has a number of advantages. First, the implementation of covert communication does not incur high resource consumption, which makes it suitable for resource-constrained devices and networks. Second, the implementation of covert communication does not rely on the execution of a complicated authentication functionality. Third, the covert communication promises a stronger protection on transmission compared to the information-theoretical secrecy approach as it can hide the transmission itself and thus the information carried by this transmission becomes immune to interception. Also, the performance of covert communication is independent of the adversary's competence.  Nevertheless, the covert communication cannot hide the presence of transmission completely and there will be a non-trivial  probability of detection by the adversary. Therefore, the following two questions arise: 1) how to reduce the detection probability and improve the communication covertness so as to achieve enhanced covert communication? 2) is there any countermeasure when the adversary attempts to wiretap the transmission once it has detected the transmission?

To answer the above questions, we resort to the friendly jamming approach, in which friendly jammers emit jamming signals to interfere with the reception at the adversary (Fig.~\ref{fig:covert_enhanced_countermeasure}(b)). In this way, the jamming signals will increase the uncertainty in the received signal at the adversary, and thereby, the detection error of the adversary will increase and the communication covertness will  improve (Fig.~\ref{fig:covert_enhanced_countermeasure}(d))~\cite{9580594}. In addition, since the jamming signals emitted by the jammers will increase the interference power received at the adversary, even in the case that the transmission has been detected, the adversary may not be able to decode the received signal successfully. This will ensure information secrecy. In this way, friendly jamming-assisted covert communication (Fig.~\ref{fig:covert_enhanced_countermeasure}(c)) will provide an additional level of protection for transmissions (Fig.~\ref{fig:covert_enhanced_countermeasure}(e)). 



\subsection{Contributions}
\label{subsec:contribution}

We propose a friendly jamming-assisted covert communication method  to secure a large-scale D2D network. As shown in Fig.~\ref{fig:system_model}(a), the network under our consideration involves: 1) a large number of power-controlled D2D transmitters cooperating with friendly jammers (which emit jamming signals), 2) their dedicated D2D receivers, and 3) spatially distributed adversaries, namely, wardens. The D2D transmitters randomly transmit information to their dedicated D2D receivers. The wardens aim to detect the D2D transmissions by following a threshold-based rule on their received signal powers, and once detect, they will wiretap the D2D transmissions. In this case, the decision-making of the wardens is after that of the D2D transmitters, which can be captured by using the framework of a two-stage Stackelberg game with the wardens as the followers at the lower stage and the D2D transmitters together with the friendly jammers as the leaders at the upper stage. The main contributions of this paper can be summarized as follows.
\begin{itemize}
\item We propose a friendly jamming-assisted covert communication approach to secure a large-scale D2D network. In this network, the friendly jammers are deployed to cooperate with the D2D network so as to jointly enhance the communication covertness and information secrecy for the D2D transmissions.

\item We model the combat between the wardens and the D2D network (i.e., the D2D transmitters and the jammers) by a two-stage Stackelberg game. At the lower stage, the wardens aim to minimize their own detection errors. At the upper stage, the D2D network jointly determines the D2D transmissions power and jamming power in order to maximize network utility subject to the constraint on communication covertness.

\item We use stochastic geometry to model the spatial configuration of the large-scale D2D network, namely, the locations of the D2D transmitters, the jammers, and the wardens, and characterize their performance metrics in order to quantitatively analyze the system-level network performance. 

\item To obtain the equilibrium of our two-stage Stackelberg game, we first analyze the characterizes of the lower-stage problem for the wardens and apply the Rosenbrock method to solve it. The solution is regarded as the best response from the lower stage. Given this best response, we develop a bi-level optimization algorithm based on SCA method to obtain the optimal strategy for the D2D network.

\item We present simulation results to verify the optimality of the obtained strategy. We also evaluate and discuss the performance of our proposed approach and demonstrate its advantages by comparing it with the conventional information-theoretical secrecy approach. 

\end{itemize}


\subsection{Organization}
The rest of the paper is organized as follows. In Section~\ref{sec:related_work}, we present the related work. In Section~\ref{sec:system_model}, we present the system model and the derivation of the performance metrics. The problem statement and the game formulation are presented in Section~\ref{subsec:game_formulation}. In Section~\ref{sec:equilibrium_game}, we develop a bi-level algorithm to solve the problem. The numerical results and discussion are presented in Section~\ref{sec:performance}, which is followed by the conclusions in Section~\ref{sec:conclusion}. The symbols used in this paper together with their descriptions and default values are given in Table~\ref{tab:notation_value}. 

\begin{table*}[!]
	\centering
	\caption{Symbols and descriptions}
	\begin{tabular}{|c|l|c|}
		\hline
		\hline
		{\bf{Symbol}} & {\bf{Description}} & {\bf{Default value}}\\
		\hline
		$d$, $\left.w\right|_d$, $j$ & \makecell[l]{Indexes of the typical D2D transmitter, the nearest warden of \\D2D transmitter $d$, and jammer, respectively.} & --- \\
        \hline
		$\cal{D}$, $\cal{W}$, $\cal{J}$ & Sets of the D2D transmitters, wardens, and jammers, respectively. & --- \\
        \hline
		\makecell[c]{$\Phi_{\cal{D}}$, $\Phi_{\cal{W}}$, $\Phi_{\cal{J}}$,\\ $\Phi_{\left\{\left.{\cal{D}}\backslash \left\{d\right\}\right| d \right\}}$,\\ $\Phi_{\left\{\left.{\cal{D}}\backslash \left\{d\right\}\right| \left.w\right|_d \right\}}$, \\$\Phi_{\left\{\left.{\cal{J}}\right| d \right\}}$, $\Phi_{\left\{\left.{\cal{J}}\right| \left.w\right|_d \right\}}$} & \makecell[l]{PP of D2D transmitters, PP of wardens, PP of \\jammers, PP of the D2D transmitters that excludes D2D \\transmitter $d$ with D2D transmitter $d$ as the observation point, \\PP of the D2D transmitters that excludes D2D transmitter $d$ \\with warden $\left.w\right|_d$ as the observation point, PP of the jammers \\with D2D transmitter $d$ as the observation point, and PP of the \\jammers with warden $\left.w\right|_d$ as the observation point, respectively.} & --- \\
		\hline
		$\lambda_{{\cal{D}}}$, $\lambda_{\cal{W}}$, $\lambda_{\cal{J}}$ & Densities of $\Phi_{{\cal{D}}}$, $\Phi_{\cal{W}}$, and $\Phi_{\cal{J}}$, respectively. & $0.1\slash {\rm{m}}^2$, $0.01\slash {\rm{m}}^2$, $0.1\slash {\rm{m}}^2$.\\	
		\hline
		${\cal{D}}_0$, ${\cal{D}}_1$ & Events that D2D transmitter is inactive and active, respectively. & --- \\	
        \hline
		${\mathbb{P}}^{{\cal{D}}_0}$, ${\mathbb{P}}^{{\cal{D}}_1}$ & \makecell[l]{Probabilities of events ${\cal{D}}_0$ and ${\cal{D}}_1$, respectively.} & $0.5$, $0.5$. \\	
		\hline
		$p^{\rm{D}}$, $p^{\rm{J}}$, $\tau$ & \makecell[l]{D2D transmission power, jamming power, and power detection \\threshold, respectively.} & \makecell[c]{$\left[0, 30\right] {\rm{dBm}}$, \\$\left[0, 30\right] {\rm{dBm}}$~\cite{8627099}, ---.} \\
		\hline
		$g_{d'd}$, $r_{d'd}$, $R$ & \makecell[l]{Channel and distance between D2D transmitter $d'$ and the \\typical D2D receiver, and distance between D2D transmitter \\and its dedicated D2D receiver, respectively.} & $\exp\left(1\right)$~\cite{haenggi2012stochastic}, ---, $1{\rm{m}}$~\cite{6609136}.\\
		\hline
		$\alpha$, $\varepsilon$, $\xi^{\rm{D}}$, $\xi^{\rm{W}}$ & \makecell[l]{Path-loss exponent, threshold of detection error, SINR threshold \\at D2D receiver, and SINR threshold at warden, respectively.} & \makecell[c]{$4$, $0.01$,\\ $-10 {\rm{dB}}$, $-10 {\rm{dB}}$~\cite{8408843}.} \\
		\hline
		$N_d$, $N_{\left.w\right|_d}$ & Noises at the typical D2D receiver and warden $\left.w\right|_d$, respectively. & $- 90{\rm{dBm}}$, $- 90{\rm{dBm}}$~\cite{8627099} \\
		\hline
		\makecell[c]{$S^{{\cal{D}}_1}_{d}$, $S^{{\cal{J}}}_{d}$, \\$S^{{\cal{D}}_1}_{\left.w\right|_d}$, $S^{{\cal{J}}}_{\left.w\right|_d}$} & \makecell[l]{Sum of signal powers from D2D transmitter $d'\in \Phi_{\left\{\left.{\cal{D}}\backslash \left\{d\right\}\right| d \right\}}$ \\received at the typical D2D receiver, sum of signal powers \\from jammer $j \in \Phi_{\left\{\left.{\cal{J}} \right| d\right\}}$ received at the typical D2D receiver, \\sum of signal powers from D2D transmitter $d' \in \Phi_{\left\{\left.{\cal{D}} \backslash \left\{d\right\}\right| \left.w\right|_d\right\}}$ \\received at warden $\left.w\right|_d$, and sum of signal powers from \\jammer $j \in \Phi_{\left\{\left.{\cal{J}} \right| \left.w\right|_d\right\}}$ received at warden $\left.w\right|_d$, respectively.} & --- \\
		\hline
		$w^{\rm{D}}$, $w^{\rm{J}}$ & \makecell[l]{Reward of stabilizing D2D link conditioned on secure \\communication and cost of unit jamming power, respectively.} & $1$, $1$ \\
		\hline
		$f_a$, $F_a$ & PDF and CDF of a random variable $a$, respectively. & --- \\
		\hline
		${\mathbb{P}}$, ${\mathbb{F}}$ & Probabilistic and expectation operators, respectively. & --- \\
		\hline
		${\mathcal{L}}$, ${\mathcal{L}}^{-1}$ & Laplace and inverse Laplace transforms, respectively. & --- \\
		\hline
	\end{tabular}
	\label{tab:notation_value}
\end{table*}


\section{Related Work}
\label{sec:related_work}

\subsection{Covert Communication}
\label{subsec:related_CC}

Covert communication promises reliable while undetectable transmission from the legitimate transmitter to its intended receiver, which thereby completely secures the legitimate transmission~\cite{forouzesh2018information}. Covert communication was adopted to secure a large-scale wireless network in~\cite{8422940}, where the uncertainty caused by the co-channel interference has been leveraged to enhance the communication covertness. Similar idea was adopted in~\cite{van2022novel},~\cite{9551936}, and~\cite{9736993}. In~\cite{van2022novel}, a novel backscatter communication was proposed, where an original message is divided into two parts (i.e., active message and backscatter message). The backscatter message is transmitted by backscattering the active signal (i.e., the signal for transmitting the active message) via ambient backscatter tag. Therein, the active signal can be used to shelter the presence of the backscatter signal (i.e., the signal for transmitting the backscatter message), which thereby can be used to carry secret information. The authors in~\cite{9551936} studied an Internet of Things (IoT) network and used the overt channel as the ``spectrum shelter'' of the covert channel. The interference from the overt channel was leveraged to hide the presence of the IoT transmission so as to enhance the network security. A large-scale IoT system involving massive IoT gateways (GWs) and IoT devices was secured in~\cite{9736993} against the adversaries from transmission detection. In the system, the in-band full-duplex (IBFD) IoT GWs emit artificial noise, which together with the co-channel interference from the IoT devices can mislead the adversaries in making their decisions. Different from the aforementioned works that leverage the uncertainty caused by the interference/noise, reconfigurable intelligent surface (RIS) was introduced in~\cite{9108996} to improve the communication covertness. Therein, by using RIS to strengthen the directionality of the intended signal, constructive and destructive effects can be achieved on the signals received by the intended receiver and the unintended receiver (i.e., an adversary), respectively. A scenario in which the adversary is unaware of the time of the communication attempt from the legitimate transmitter was investigated in~\cite{7579596}. In this scenario, the legitimate transmitter selects one single slot out of multiple slots for transmission. An upper bound of the achievable covert rate was analytically derived as a performance metric to evaluate the effectiveness of covert communication.


\subsection{Friendly Jamming}
\label{subsec:related_friendly_jamming}

The main idea of friendly jamming is to emit jamming signal that will interfere with the receptions at the eavesdropper to achieve  confidentiality of legitimate transmissions. Thus the jamming signals  increase the decoding difficulty for the eavesdropper and  improve the secrecy outage probability, or equivalently, the secrecy rate~\cite{5401476, 8647745}. Friendly jamming was adopted together with bandwidth allocation in~\cite{9162671} to secure the unmanned aerial vehicle (UAV)-enabled wireless communication. A UAV-enabled friendly jamming scheme was proposed in~\cite{8796506} to secure industrial IoT (IIoT) against eavesdropping, and its effectiveness in reducing the secrecy outage probability was verified both analytically and numerically. Except for jamming the wiretap channel so as to reduce the secrecy outage probability, friendly jamming can also be applied to improve the communication covertness. This is due to the fact that the jamming signal emitted by the friendly jammer can also increase the uncertainty in the received signal power of the adversary  and thereby increase the detection error. For example, the authors in~\cite{9580594} deployed a friendly jammer to transmit jamming signal and prevent the adversary from detecting transmission of the local model updates in order to achieve a covert federated learning (FL) process. As the jamming signal can also reduce transmission reliability, an optimization problem was formulated to determine the optimal jamming power such that both the link reliability and communication covertness for the local model updates in the FL process can be guaranteed. \cite{9361424} considered a scenario where a transmitter intends to communicate with multiple receivers via orthogonal frequency bands and the adversary is able to monitor all the bands aiming at detecting the transmissions. A friendly jammer was deployed to broadcast jamming signals to distort the adversary's observation in order to increase the detection error for the adversary over all the bands. Different from other works, the authors in~\cite{9390411} considered a scenario where the adversary can optimize not only its detection threshold but also its location aiming to minimize the covert outage probability. To defend against such a powerful adversary, a full-duplex transceiver was deployed, and the jamming signal transmitted by this transceiver was leveraged to maximize the throughput of the legitimate user subject to the constraint on communication covertness.



\section{System Model, Assumptions, and Performance Metrics}
\label{sec:system_model}

\subsection{Network Model}
\label{subsec:network_model}

\begin{figure}[!]
	\centering
	\includegraphics[width=1\textwidth,trim=230 230 145 80, clip]{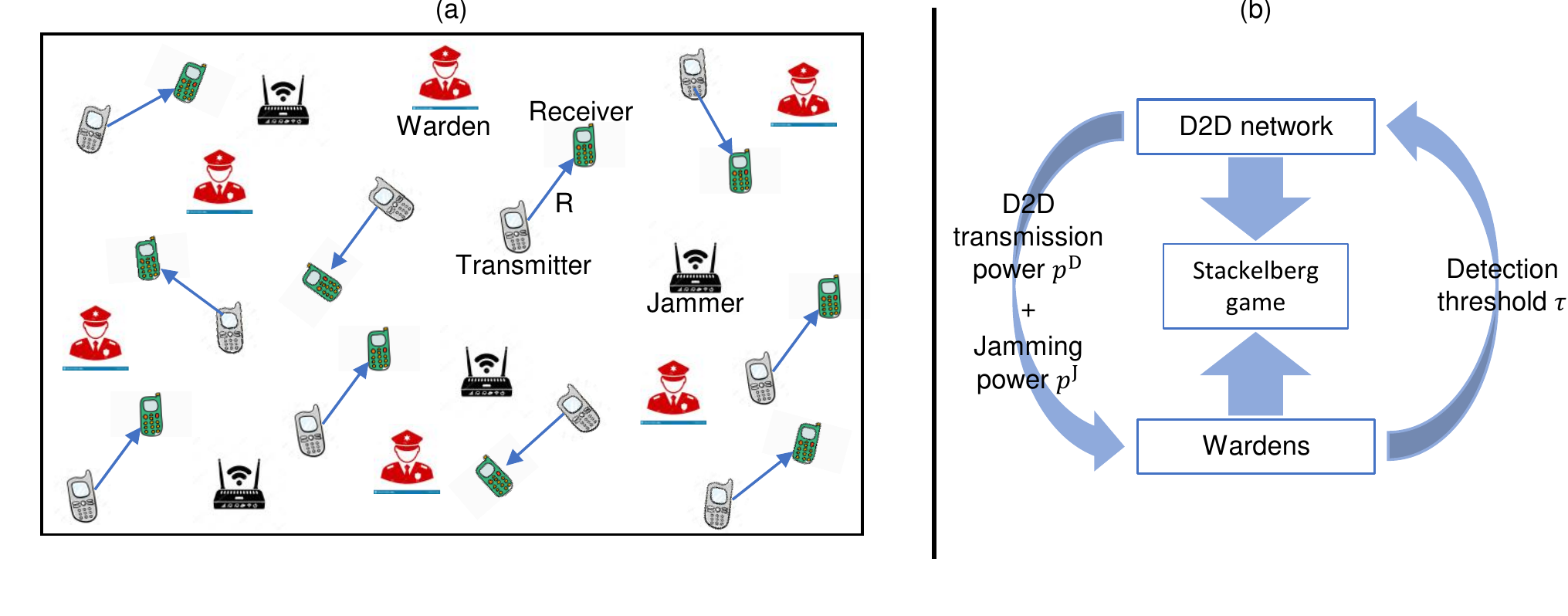}
	\caption{(a) D2D transmitters, D2D receivers, friendly jammers, and the wardens; (b) Stackelberg game-based problem formulation with the wardens at the lower stage and the D2D network (i.e., D2D transmitters and jammers) at the upper stage.}
	\label{fig:system_model}
\end{figure}

We consider a large-scale D2D network under the threat of wardens (Fig.~\ref{fig:system_model}(a)). In the network, the spatial distribution of the D2D transmitters follows a homogeneous Poisson point process (PPP) $\Phi_{\cal{D}}$ with density $\lambda_{\cal{D}}$, and each of which has a dedicated D2D receiver located at distance $R$ in a random orientation. The D2D transmitters and their dedicated D2D receivers form a Poisson bipolar network~\cite{haenggi2012stochastic}. The ALOHA channel access scheme is adopted for the D2D network, where each D2D transmitter becomes active independently. This event is denoted by ${\cal{D}}_1$ and it occurs with probability ${\mathbb{P}}^{{\cal{D}}_1}$~\cite{7345601}. Accordingly, each D2D transmitter will independently become inactive with probability ${\mathbb{P}}^{{\cal{D}}_0} \triangleq 1 - {\mathbb{P}}^{{\cal{D}}_1}$, where ${\cal{D}}_0$ denotes the corresponding inactive event. To analyze the performance of the D2D network, we condition on that there is a D2D receiver at the origin which is considered to be the typical D2D receiver, and its associating D2D transmitter is the typical D2D transmitter and denoted by D2D transmitter $d$.  Following~\cite{7464352}, we consider that the wardens are spatially distributed by following an independent homogeneous PPP $\Phi_{\cal{W}}$ with density $\lambda_{\cal{W}}$. Also, the spatial distribution of the jammers follows another independent homogeneous PPP $\Phi_{\cal{J}}$ with density $\lambda_{\cal{J}}$. 

For D2D transmitter $d$, since the nearest warden, denoted by $\left.w\right|_d$, is the most threatening warden, we focus on a  scenario where D2D transmitter $d$ defends against warden $\left.w\right|_d$.  Accordingly, we consider that warden $\left.w\right|_d$ aims to detect, and once detect, wiretaps the transmission of D2D transmitter $d$. All the entities in the network are equipped with a single antenna and all the channels are assumed to experience Rayleigh fading with unit mean~\cite{haenggi2012stochastic}. Note that similar to~\cite{7056528}, the discussion in the rest of this paper is based on the performance of the representative network nodes (i.e., D2D transmitter $d$, the typical D2D receiver, and warden $\left.w\right|_d$).

\subsection{Performance Metrics}

\subsubsection{Successful Transmission Probability for the D2D Transmitter}

The received signal power of the typical D2D receiver regarding the activation status of its associating D2D transmitter (i.e., D2D transmitter $d$) is
\begin{equation}
y_d = \left\{
\begin{aligned}
& p^{\rm{D}} g_{d} R^{-\alpha} + S_{d}^{{\cal{D}}_1} + S_{d}^{{\cal{J}}} + N_d, & {\text{if}} \, {\cal{D}}_1, \\
& S_{d}^{{\cal{D}}_1} + S_{d}^{{\cal{J}}} + N_d, & {\text{if}} \, {\cal{D}}_0,  
\end{aligned} \right.
\end{equation}
where $S_d^{{\cal{D}}_1} = p^{\rm{D}} \sum_{d' \in \Phi_{\left\{\left.{\cal{D}} \backslash \left\{d\right\}\right| d\right\}}} {\mathbbm{1}}_{d'} g_{d'd} r_{d'd}^{-\alpha}$ and $S_d^{{\cal{J}}} = p^{\rm{J}} \sum_{j \in \Phi_{\left\{\left.{\cal{J}} \right| d \right\}}} g_{j d} r_{j d}^{-\alpha}$ are the sum of the signal powers from D2D transmitter $d' \in \Phi_{\left\{\left.{\cal{D}} \backslash \left\{d\right\}\right| d \right\}}$ and that from jammer $j \in \Phi_{\left\{\left.{\cal{J}} \right| d\right\}}$ received at the typical D2D receiver, respectively, $g_d$ is the channel between D2D transmitter $d$ and the typical D2D receiver, and $N_d$ is the noise power at the typical D2D receiver. Therein, $\Phi_{\left\{\left.{\cal{D}}\backslash \left\{d\right\}\right| d \right\}}$ and $\Phi_{\left\{\left.{\cal{J}} \right| d\right\}}$ are the point process (PP) of the D2D transmitters that excludes D2D transmitter $d$ with the typical D2D receiver as the observation point and that of the jammers with the typical D2D receiver as the observation point, respectively, ${\mathbbm{1}}_{d'}$ is the activation indicator of D2D transmitter $d' \in \Phi_{\left\{\left.{\cal{D}} \backslash \left\{d\right\}\right| d\right\}}$ and equal to $1$ if D2D transmitter $d'$ is active and $0$ otherwise, $p^{\rm{D}}$ and $p^{\rm{J}}$ are the D2D transmission power and jamming power, respectively,  $g_{d'd}$ and $r_{d'd}$ are the channel and distance between D2D transmitter $d' \in \Phi_{\left\{\left.{\cal{D}} \backslash \left\{d\right\}\right| d\right\}}$ and the typical D2D receiver, respectively, and $g_{jd}$ and $r_{jd}$ are the channel and distance between jammer $j \in \Phi_{\left\{\left.{\cal{J}} \right| d \right\}}$ and the typical D2D receiver, respectively. Accordingly, the SINR at the typical D2D receiver regarding the activation status of its associating D2D transmitter (i.e., D2D transmitter $d$) is
\begin{equation}\label{eq:SINR}
{\rm{SINR}}_d \left(p^{\rm{D}}, p^{\rm{J}}\right) = \left\{\begin{aligned}
& \frac{p^{\rm{D}} g_d R^{-\alpha}} {S_{d}^{{\cal{D}}_1} + S_{d}^{{\cal{J}}} + N_d}, & {\text{if}} \, {\cal{D}}_1,\\
& 0, & {\text{if}} \, {\cal{D}}_0.\\
\end{aligned}\right.
\end{equation}

\begin{figure}[!]
	\centering
	\includegraphics[width=0.5\textwidth,trim=10 10 10 70, clip]{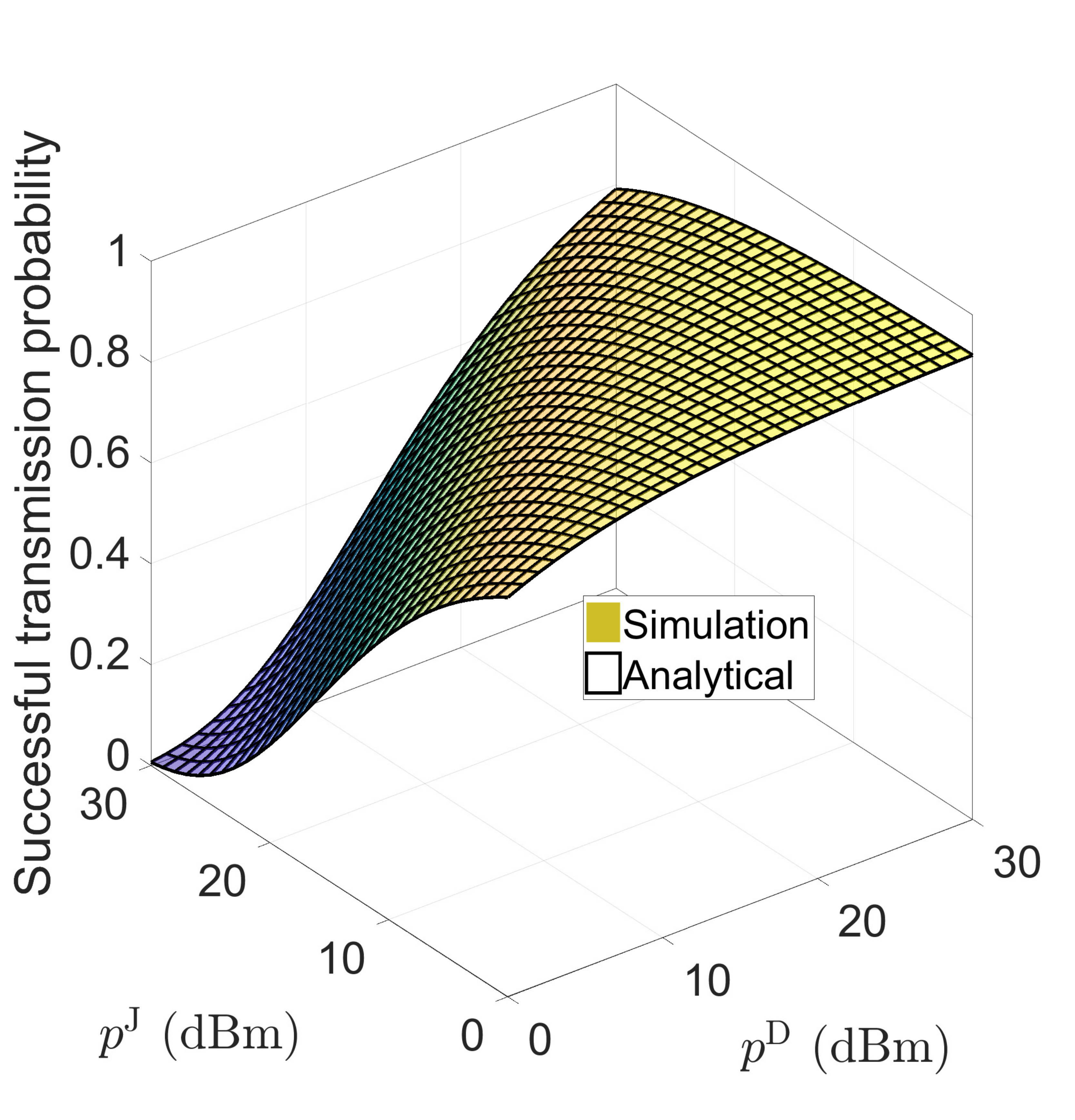}
	\caption{Comparison between simulation results and analytical results on successful transmission probability.}
	\label{fig:simulation_vs_analysis_SINR}
\end{figure}

Denoting the SINR threshold at the D2D receiver by $\xi^{\rm{D}}$ and conditioned on the active status of D2D transmitter $d$ (i.e., ${\cal{D}}_1$), we have the successful transmission probability for the D2D transmitter $d$ as follows:
\begin{equation}\label{eq:ProbRel_SINR}
{\mathbb{P}} \left[\left.{\rm{SINR}}_d \left(p^{\rm{D}}, p^{\rm{J}}\right) > \xi^{\rm{D}} \right| {\cal{D}}_1 \right].
\end{equation}
The detailed derivation of~(\ref{eq:ProbRel_SINR}) is given in \textbf{Appendix~\ref{app:SINR_d2d}}. We validate the derivation of~(\ref{eq:ProbRel_SINR}) by comparing the analytical results with the simulation results in Fig.~\ref{fig:simulation_vs_analysis_SINR}. Note that the simulation results are generated by using Monte Carlo method using the parameter setting given in Table~\ref{tab:notation_value}. We assume the same parameter setting  in the rest of this paper. It can be observed from Fig.~\ref{fig:simulation_vs_analysis_SINR} that our analytical results match with the simulation results.

\subsubsection{Detection Error at the Warden}

The wardens receive the signals transmitted from the D2D transmitters and jammers and aim to detect the D2D transmission independently. Accordingly, for warden $\left.w\right|_d \in \Phi_{\cal{W}}$, its received signal power regarding the activation status of its target D2D transmitter (i.e., D2D transmitter $d$) can be expressed as follows:
\begin{equation}\label{eq:received_signal_power_w|d}
y_{\left.w\right|_d} = \left\{
\begin{aligned}
& p^{\rm{D}} g_{d\left.w\right|_d} r_{d\left.w\right|_d}^{-\alpha}+ S_{\left.w\right|_d}^{{\cal{D}}_1} + S_{\left.w\right|_d}^{{\cal{J}}} + N_{\left.w\right|_d}, & {\text{if}} \, {\cal{D}}_1, \\
& S_{\left.w\right|_d}^{{\cal{D}}_1} + S_{\left.w\right|_d}^{{\cal{J}}} + N_{\left.w\right|_d}, & {\text{if}} \, {\cal{D}}_0,  
\end{aligned} \right.
\end{equation}
where $S_{\left.w\right|_d}^{{\cal{D}}_1} = p^{\rm{D}} \sum_{d' \in \Phi_{\left\{\left.{\cal{D}} \backslash \left\{d\right\}\right| \left.w\right|_d\right\}}} {\mathbbm{1}}_{d'} g_{d'\left.w\right|_d} r_{d'\left.w\right|_d}^{-\alpha}$ and $S_{\left.w\right|_d}^{{\cal{J}}} = p^{\rm{J}} \sum_{j \in \Phi_{\left\{\left.{\cal{J}} \right| \left.w\right|_d\right\}}} g_{j\left.w\right|_d} r_{j\left.w\right|_d}^{-\alpha}$ are the sum of the signal powers from D2D transmitter $d' \in \Phi_{\left\{\left.{\cal{D}} \backslash \left\{d\right\}\right| \left.w\right|_d\right\}}$ and that from jammer $j \in \Phi_{\left\{\left.{\cal{J}} \right| \left.w\right|_d\right\}}$ received at warden $\left.w\right|_d$, respectively, $g_{d \left.w\right|_d}$ and $r_{d \left.w\right|_d}$ are the channel and distance between D2D transmitter $d$ and warden $\left.w\right|_d$, respectively, and $N_{\left.w\right|_d}$ is the additive noise at warden $\left.w\right|_d$. Therein, $\Phi_{\left\{\left.{\cal{D}}\backslash \left\{d\right\}\right| \left.w\right|_d \right\}}$ and $\Phi_{\left\{\left.{\cal{J}} \right| \left.w\right|_d\right\}}$ are the PP of the D2D transmitters that excludes D2D transmitter $d$ with warden $\left.w\right|_d$ as the observation point and that of the jammers with warden $\left.w\right|_d$ as the observation point, respectively, $g_{d' \left.w\right|_d}$ and $r_{d' \left.w\right|_d}$ are the channel and distance between D2D transmitter $d' \in \Phi_{\left\{\left.{\cal{D}} \backslash \left\{d\right\}\right| \left.w\right|_d\right\}}$ and warden $\left.w\right|_d$, respectively, and $g_{j \left.w\right|_d}$ and $r_{j \left.w\right|_d}$ are the channel and distance between jammer $j \in \Phi_{\left\{\left.{\cal{J}} \right| \left.w\right|_d\right\}}$ and warden $\left.w\right|_d$, respectively. 

The warden uses a threshold-based detection rule to detect the transmission of its target D2D transmitter~\cite{9108996}. Specifically, the warden will assume that its target D2D transmitter is active if its received signal power is larger than a predetermined detection threshold, denoted by $\tau$. Otherwise, it will assume that its target D2D transmitter is inactive. It will be a false alarm (FA) for warden  $\left.w\right|_d$ if its received signal power is larger than the predetermined detection threshold $\tau$ (i.e., $y_{\left.w\right|_d} > \tau$) while D2D transmitter $d$ is inactive (i.e., ${\cal{D}}_0$). On the other hand, it will be a miss detection (MD) for warden $\left.w\right|_d$, if its received signal power is lower than the predetermined detection threshold $\tau$ (i.e., $y_{\left.w\right|_d} < \tau$) while D2D transmitter $d$ is active (i.e., ${\cal{D}}_1$). 

The FA and MD probabilities for warden $\left.w\right|_d \in \Phi_{\cal{W}}$ are defined as
\begin{equation}\label{eq:ProbFA}
{\mathbb{P}}^{\text{FA}}_{\left.w\right|_d} \left(p^{\rm{D}}, p^{\rm{J}}, \tau\right) = {\mathbb{P}} \left[\left. y_{\left.w\right|_d} > \tau \right| {\cal{D}}_0 \right] = {\mathbb{P}} \left[S_{\left.w\right|_d}^{{\cal{D}}_1} + S_{\left.w\right|_d}^{{\cal{J}}} + N_{\left.w\right|_d} > \tau \right]
\end{equation}
and 
\begin{equation}\label{eq:ProbMD}
{\mathbb{P}}^{\rm{MD}}_{\left.w\right|_d} \left(p^{\rm{D}}, p^{\rm{J}}, \tau\right) = {\mathbb{P}} \left[\left. y_{\left.w\right|_d} < \tau \right| {\cal{D}}_1 \right] = {\mathbb{P}} \left[ p^{\rm{D}} g_{d\left.w\right|_d} r_{d\left.w\right|_d}^{-\alpha} + S_{\left.w\right|_d}^{{\cal{D}}_1} + S_{\left.w\right|_d}^{{\cal{J}}} + N_{\left.w\right|_d} < \tau \right],
\end{equation}
respectively, and the specific expressions of which are derived in \textbf{Appendices~\ref{app:ProbFA}} and~\textbf{\ref{app:ProbMD}}, respectively. Again, we use simulation results to verify the derivations of ${\mathbb{P}}^{\text{FA}}_{\left.w\right|_d} \left(p^{\rm{D}}, p^{\rm{J}}, \tau\right)$ and ${\mathbb{P}}^{\text{MD}}_{\left.w\right|_d} \left(p^{\rm{D}}, p^{\rm{J}}, \tau\right)$. 

\begin{figure}
     \centering
     \begin{minipage}{8cm}
		\centering
		\includegraphics[width=1\textwidth,trim=10 0 15 10,clip]{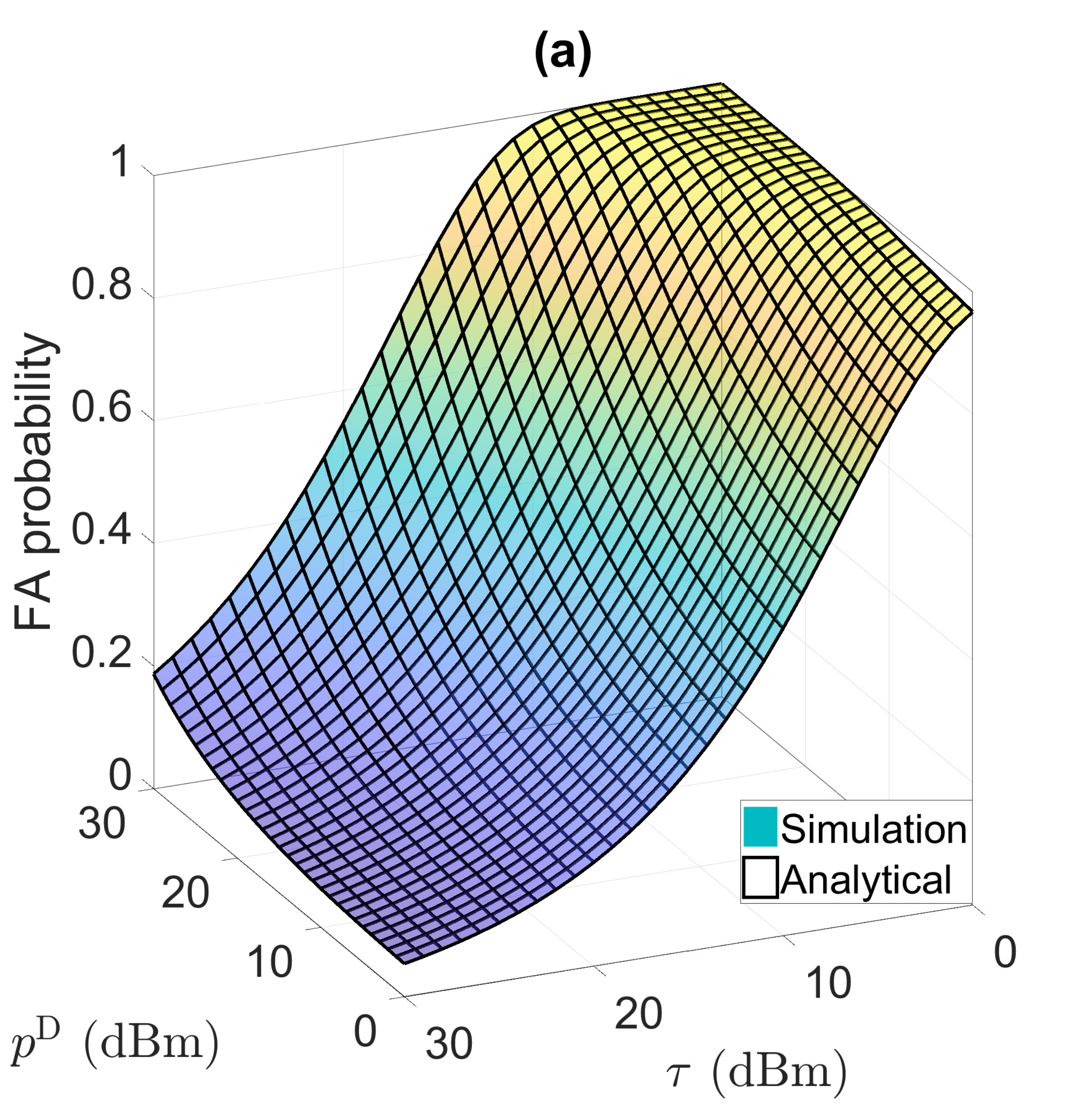}
     \end{minipage}
     \begin{minipage}{8cm}
		\centering
		\includegraphics[width=1\textwidth,trim=10 0 15 10,clip]{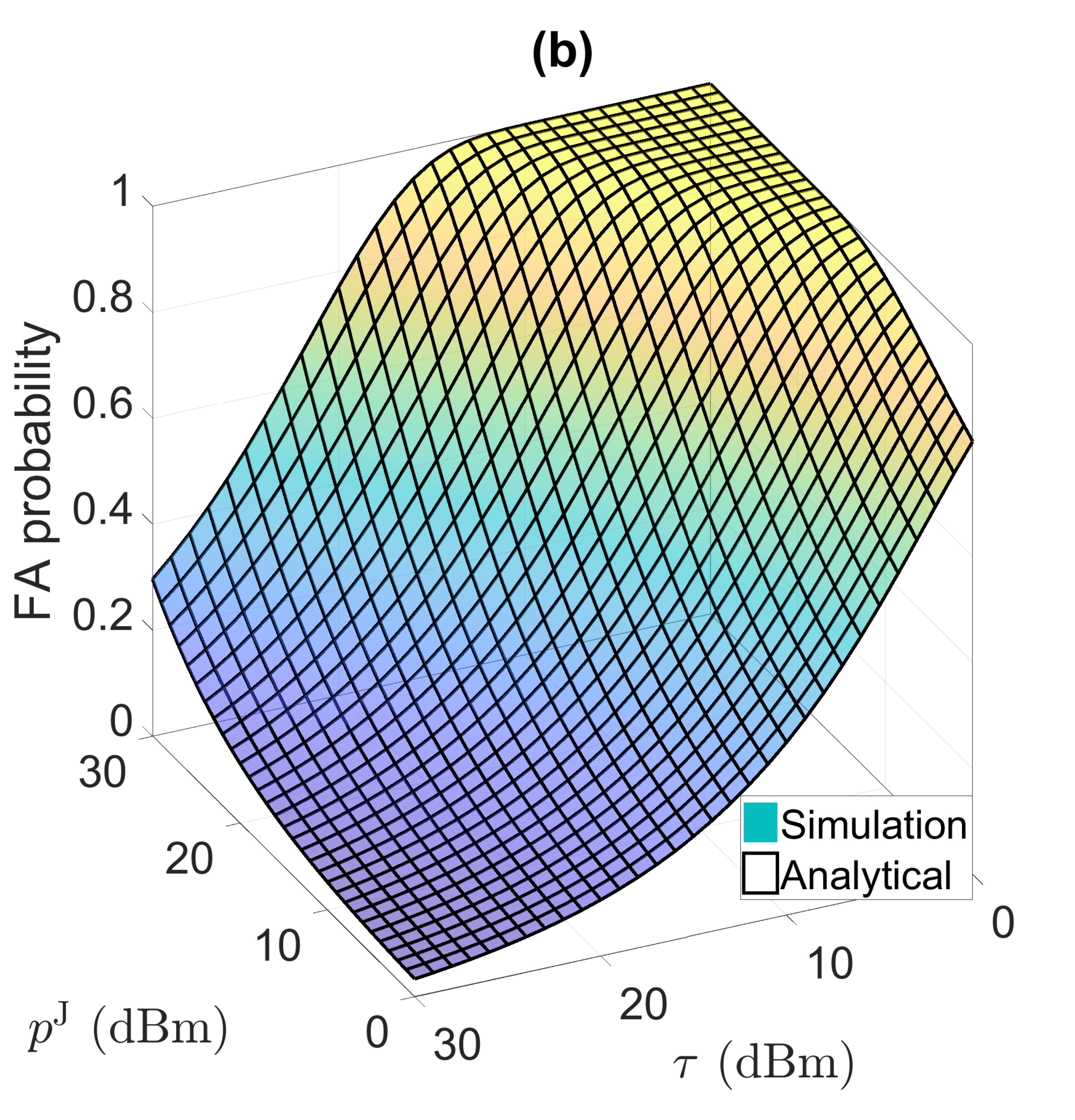}
     \end{minipage}
        \caption{Comparison between the simulation results and analytical results on FA probability with (a) $p^{\rm{J}} = 15 {\rm{dBm}}$ and (b) $p^{\rm{D}} = 15 {\rm{dBm}}$.}
        \label{fig:simulation_vs_analysis_FA}
\end{figure}

\begin{figure}
     \centering
     \begin{minipage}{8cm}
		\centering
		\includegraphics[width=1\textwidth,trim=10 0 30 10,clip]{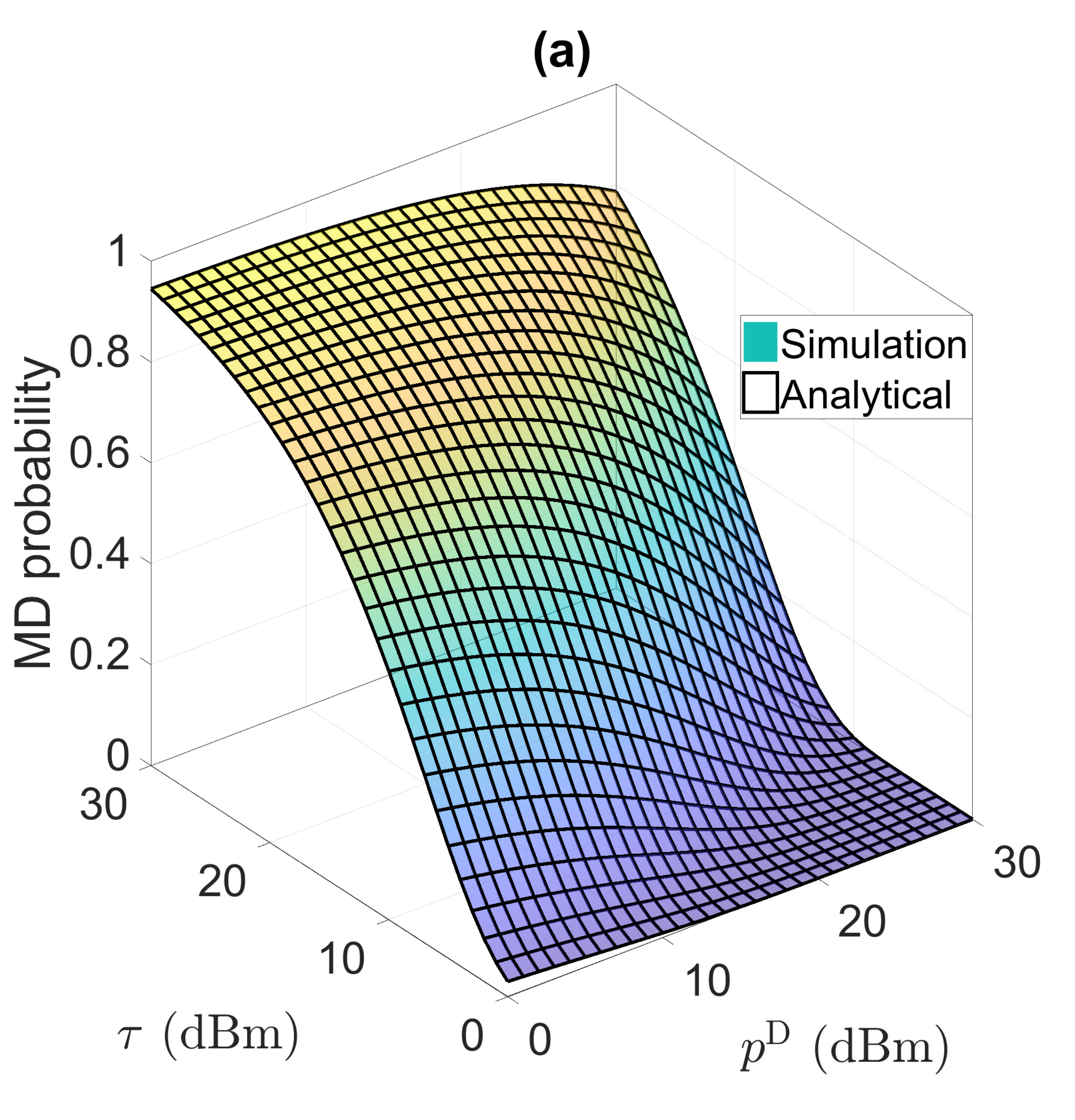}
     \end{minipage}
     \begin{minipage}{8cm}
		\centering
		\includegraphics[width=1\textwidth,trim=10 0 30 10,clip]{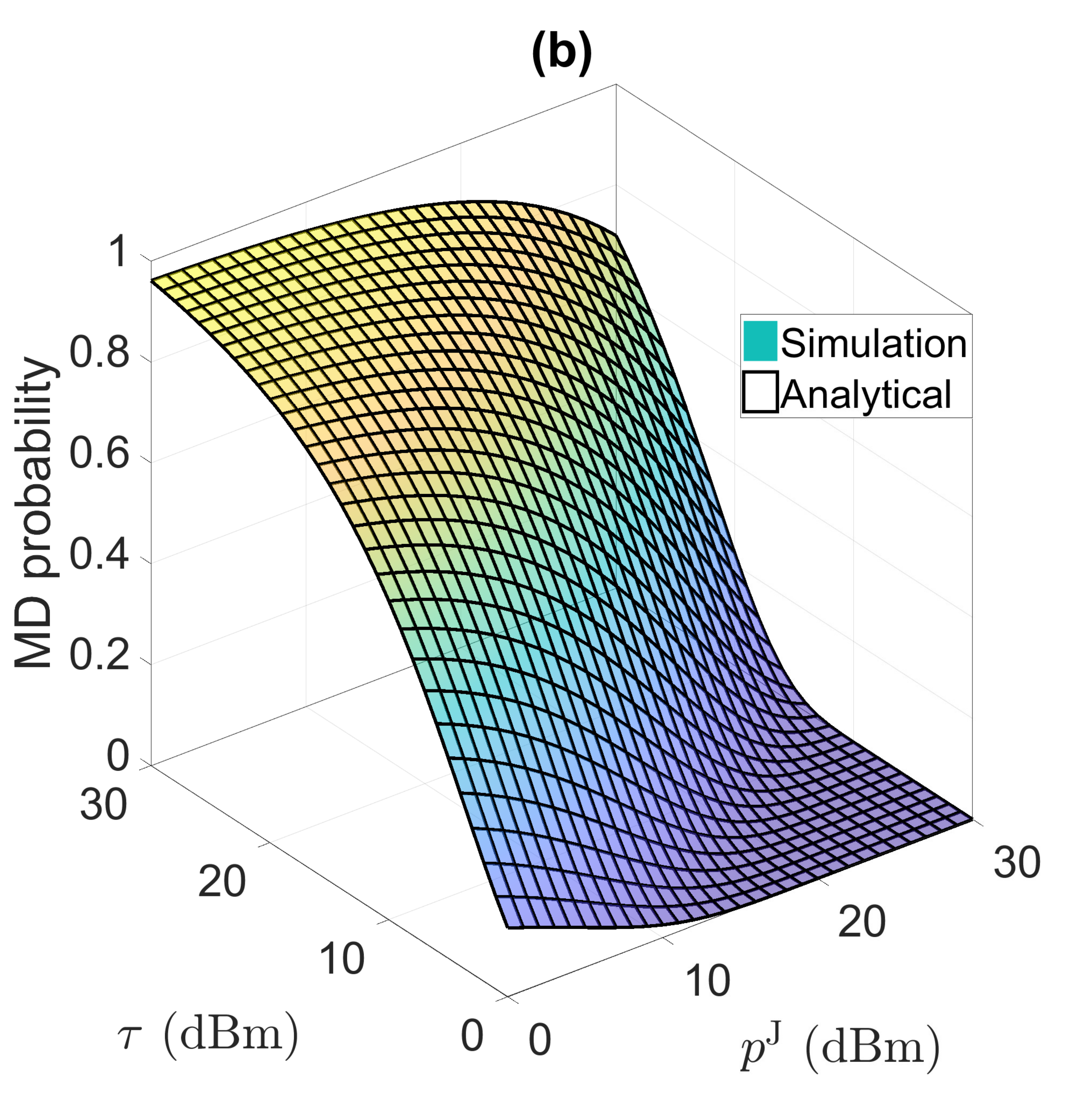}
     \end{minipage}
        \caption{Comparison between the simulation results and analytical results on MD probability with (a) $p^{\rm{J}} = 15 {\rm{dBm}}$ and (b) $p^{\rm{D}} = 15 {\rm{dBm}}$.}
        \label{fig:simulation_vs_analysis_MD}
\end{figure}

As shown in Fig.~\ref{fig:simulation_vs_analysis_FA}(a), we first fix $p^{\rm{J}}$ to be $15$dBm and vary the values of $p^{\rm{D}}$ and $\tau$ to compare the simulation results and analytical results on ${\mathbb{P}}^{\text{FA}}_{\left.w\right|_d} \left(p^{\rm{D}}, p^{\rm{J}}, \tau\right)$ and observe that these two results are consistent, which validates our derivation of ${\mathbb{P}}^{\text{FA}}_{\left.w\right|_d} \left(p^{\rm{D}}, p^{\rm{J}}, \tau\right)$. The same conclusion can be drawn based on the results in Fig.~\ref{fig:simulation_vs_analysis_FA}(b), which compares the simulation results and analytical results on ${\mathbb{P}}^{\text{FA}}_{\left.w\right|_d} \left(p^{\rm{D}}, p^{\rm{J}}, \tau\right)$ by fixing $p^{\rm{D}}$ to be $15$dBm and varying the values of $p^{\rm{J}}$ and $\tau$. Similar outcomes can be observed in Fig.~\ref{fig:simulation_vs_analysis_MD}, which validates our derivation of ${\mathbb{P}}^{\rm{MD}}_{\left.w\right|_d} \left(p^{\rm{D}}, p^{\rm{J}}, \tau\right)$. 

\subsubsection{Secrecy Outage Probability for the D2D Transmitter}

\begin{figure}[!]
	\centering
	\includegraphics[width=0.5\textwidth,trim=10 10 10 70, clip]{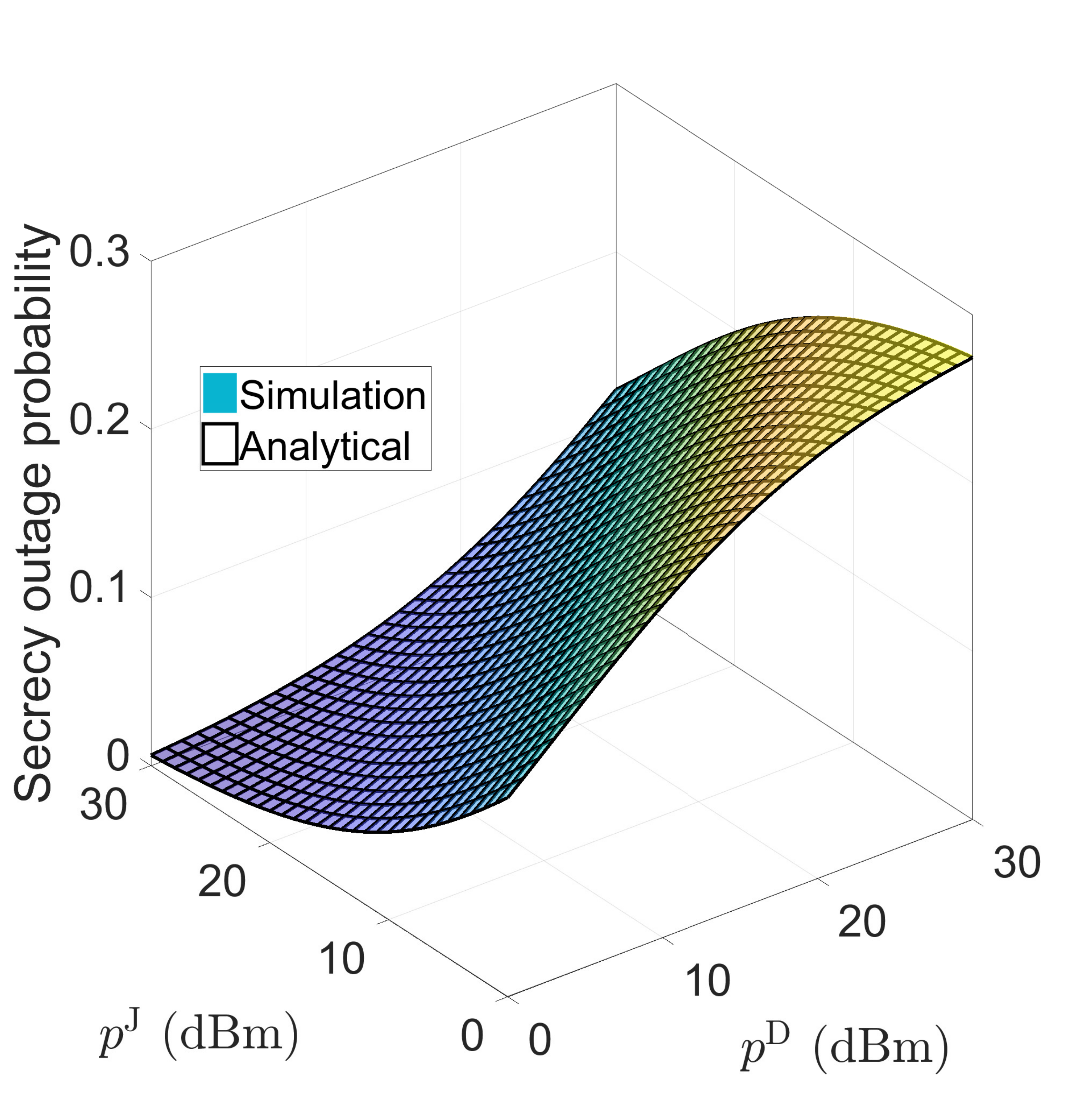}
	\caption{Comparison between simulation results and analytical results on secrecy outage probability.}
	\label{fig:simulation_vs_analysis_SecrecyOutage}
\end{figure}

According to the received signal power of warden $\left.w\right|_d$ given in~(\ref{eq:received_signal_power_w|d}) and conditioned on the active status of its target D2D transmitter (i.e., D2D transmitter $d$), the secrecy outage probability for D2D transmitter $d$ is 
\begin{equation}\label{eq:Prob_SecrecyOutage}
{\mathbb{P}} \left[\left.{\rm{SINR}}_{\left.w\right|_d} \left(p^{\rm{D}}, p^{\rm{J}}\right) > \xi^{\rm{W}} \right| {\cal{D}}_1 \right],
\end{equation}
where $\xi^{\rm{W}}$ is the SINR threshold at warden for successfully decoding the received signal and the expression of which is derived in \textbf{Appendix~\ref{app:Secrecy_outage_d2d}}. The consistency between the analytical results and simulation results on the secrecy outage probability is shown in Fig.~\ref{fig:simulation_vs_analysis_SecrecyOutage}.

\section{Problem Statement and Game Formulation}
\label{subsec:game_formulation}

We model the combat between wardens and the D2D network including D2D transmitters and the jammers in the framework of a two-stage Stackelberg game as shown in Fig.~\ref{fig:system_model}(b). Therein, the D2D network jointly decides the D2D transmission power and jamming power first, and the wardens make decision based on their received signal powers later. Hence, the wardens are the followers at the lower stage and the D2D network is the leader at the upper stage. The formulation of the two-stage Stackelberg game regarding the representative network nodes mentioned in Section~\ref{sec:system_model} is a single-leader-single-follower game and defined as follows:
\begin{itemize}
\item At the lower stage, given the strategies of the D2D network (i.e., the D2D transmission power $p^{\rm{D}}$ and jamming power $p^{\rm{J}}$) from the upper stage, warden $\left.w\right|_d$ acts as a single follower and makes decision on the detection threshold $\tau$ in order to minimize its detection error~\cite{9108996} as follows:
\begin{equation}\label{eq:warden_w_problem}
\tau^* = \arg \min_{\tau}  {\mathbb{P}}^{\text{FA}}_{\left.w\right|_d} \left(p^{\rm{D}}, p^{\rm{J}}, \tau\right) + {\mathbb{P}}^{\rm{MD}}_{\left.w\right|_d} \left(p^{\rm{D}}, p^{\rm{J}}, \tau\right),
\end{equation}
where ${\mathbb{P}}^{\text{FA}}_{\left.w\right|_d} \left(p^{\rm{D}}, p^{\rm{J}}, \tau\right)$ and ${\mathbb{P}}^{\rm{MD}}_{\left.w\right|_d} \left(p^{\rm{D}}, p^{\rm{J}}, \tau\right)$ are defined in~(\ref{eq:ProbFA}) and~(\ref{eq:ProbMD}), respectively. 

In the above, the reason that we omit ${\mathbb{P}}^{{\cal{D}}_1}$ and ${\mathbb{P}}^{{\cal{D}}_0}$ in the detection error can be explained as follows. First, as warden $\left.w\right|_d$ is unaware of the activation status of its target D2D transmitter (i.e., ${\cal{D}}_1$ and ${\cal{D}}_0$), we relax its detection error probability as follows:
\begin{equation}\label{eq:detection_prob_UB}
\begin{aligned}
& {\mathbb{P}}^{{\cal{D}}_0}{\mathbb{P}}^{\text{FA}}_{\left.w\right|_d} \left(p^{\rm{D}}, p^{\rm{J}}, \tau\right) + {\mathbb{P}}^{{\cal{D}}_1}{\mathbb{P}}^{\rm{MD}}_{\left.w\right|_d} \left(p^{\rm{D}}, p^{\rm{J}}, \tau\right)\\
\le & \max \left\{{\mathbb{P}}^{{\cal{D}}_0}, {\mathbb{P}}^{{\cal{D}}_1}\right\}\left[{\mathbb{P}}^{\text{FA}}_{\left.w\right|_d} \left(p^{\rm{D}}, p^{\rm{J}}, \tau\right) + {\mathbb{P}}^{\rm{MD}}_{\left.w\right|_d} \left(p^{\rm{D}}, p^{\rm{J}}, \tau\right)\right],
\end{aligned}
\end{equation}
where the second row of~(\ref{eq:detection_prob_UB}) is the upper bound of the detection error probability (i.e., the first row of~(\ref{eq:detection_prob_UB})) and can be further simplified as the objective shown in~(\ref{eq:warden_w_problem}). As such, the probabilistic terms related to the activation status of the D2D transmitter are avoided in the warden's objective. Also, this enables us to treat FA and MD equivalently so as to avoid any bias between them~\cite{6692447, 5486887}. 

\item At the upper stage, D2D transmitter $d$ together with the jammers work as a single leader and aim to maximize the network utility  (i.e.,~(\ref{eq:D2D_problem_obj})) subject to the constraint on communication covertness with respect to (w.r.t.) warden $\left.w\right|_d$ (i.e.,~(\ref{eq:D2D_problem_constr})). This can be expressed as a nonlinear constrained optimization problem as follows:
\begin{subequations}\label{eq:D2D_problem}
\begin{align}
&\begin{aligned}
\max_{p^{\rm{D}}, p^{\rm{J}}} & \, w^{\rm{D}} {\mathbb{P}} \left[\left.{\rm{SINR}}_d \left(p^{\rm{D}}, p^{\rm{J}}\right) > \xi^{\rm{D}} \right| {\cal{D}}_1 \right] \left\{ {\mathbb{P}}^{\rm{MD}}_{\left.w\right|_d} \left(p^{\rm{D}}, p^{\rm{J}}, \tau^\star\right) \right.\\
& \, \left. + \left[1 - {\mathbb{P}}^{\rm{MD}}_{\left.w\right|_d} \left(p^{\rm{D}}, p^{\rm{J}}, \tau^\star\right)\right] \left[1 - {\mathbb{P}} \left[\left.{\rm{SINR}}_{\left.w\right|_d} \left(p^{\rm{D}}, p^{\rm{J}}\right) > \xi^{\rm{W}} \right| {\cal{D}}_1 \right]\right]\right\} - \frac{\lambda_{\cal{J}}}{\lambda_{\cal{D}}}w^{\rm{J}} p^{\rm{J}}
\end{aligned} \label{eq:D2D_problem_obj}\\
& \quad {\text{s.t.}} \, {\mathbb{P}}^{\text{FA}}_{\left.w\right|_d} \left(p^{\rm{D}}, p^{\rm{J}}, \tau^\star\right) + {\mathbb{P}}^{\rm{MD}}_{\left.w\right|_d} \left(p^{\rm{D}}, p^{\rm{J}}, \tau^\star\right) \ge 1 - \varepsilon \label{eq:D2D_problem_constr}\\
& \quad \, \quad \, p^{\rm{D}} \in \left[{\underline{p}}^{\rm{D}}, {\overline{p}}^{\rm{D}}\right], \, p^{\rm{J}} \in \left[{\underline{p}}^{\rm{J}}, {\overline{p}}^{\rm{J}}\right], \label{eq:D2D_problem_domain_def}
\end{align}
\end{subequations}
where ${\underline{p}}^{\rm{D}}$ (${\underline{p}}^{\rm{J}}$) and ${\overline{p}}^{\rm{D}}$ (${\overline{p}}^{\rm{J}}$) are the lower bound and upper bound of the D2D transmission (jamming) power, respectively. Therein, $w^{\rm{D}}$ is the reward for guaranteeing D2D link reliability (i.e., ${\rm{SINR}}_d \left(p^{\rm{D}}, p^{\rm{J}}\right) > \xi^{\rm{D}}$) conditioned on secure communication. The secure communication happens if and only if the malicious attempt of warden $\left.w\right|_d$ is unsuccessful (i.e., miss detect the D2D transmission or secrecy outage is avoided if the D2D transmission is detected) when its target D2D transmitter (i.e., D2D transmitter $d$) is active. In this case, the secure communication excludes the FA, and the probability of which is defined as the sum of the probability that MD occurs at warden $\left.w\right|_d$ (i.e., ${\mathbb{P}}^{\rm{MD}}_{\left.w\right|_d} \left(p^{\rm{D}}, p^{\rm{J}}, \tau^\star\right)$) and the probability that secrecy outage is avoided for D2D transmitter $d$ (i.e., $1 - {\mathbb{P}} \left[\left.{\rm{SINR}}_{\left.w\right|_d} \left(p^{\rm{D}}, p^{\rm{J}}\right) > \xi^{\rm{W}} \right| {\cal{D}}_1 \right]$) in the case that its transmission is detected by warden $\left.w\right|_d$ (i.e., $1 - {\mathbb{P}}^{\rm{MD}}_{\left.w\right|_d} \left(p^{\rm{D}}, p^{\rm{J}}, \tau^\star\right)$). Hence, the probability of secure communication is given by 
\begin{equation}\label{eq:prob_secure_communication}
{\mathbb{P}}^{\rm{MD}}_{\left.w\right|_d} \left(p^{\rm{D}}, p^{\rm{J}}, \tau^\star\right) + \left[1 - {\mathbb{P}}^{\rm{MD}}_{\left.w\right|_d} \left(p^{\rm{D}}, p^{\rm{J}}, \tau^\star\right)\right] \left[1 - {\mathbb{P}} \left[\left.{\rm{SINR}}_{\left.w\right|_d} \left(p^{\rm{D}}, p^{\rm{J}}\right) > \xi^{\rm{W}} \right| {\cal{D}}_1 \right]\right].
\end{equation}

In the objective of the optimization problem (i.e.,~(\ref{eq:D2D_problem_obj})), the term $ \frac{\lambda_{\cal{J}}}{\lambda_{\cal{D}}}w^{\rm{J}} p^{\rm{J}}$ is the cost incurred by the assistance from jammers with $w^{\rm{J}}$ being the cost of unit jamming power. Here, if the area of the region occupied by the large-scale D2D network is $A$, the expected number of jammers in this region is $\lambda_{\cal{W}}A$. In this case, the total jamming power cost is $\lambda_{\cal{W}}A w^{\rm{J}} p^{\rm{J}}$. Let D2D transmitters equally share the total jamming power cost, $\frac{\lambda_{\cal{W}}A w^{\rm{J}} p^{\rm{J}}}{\lambda_{\cal{D}}A} = \frac{\lambda_{\cal{J}}}{\lambda_{\cal{D}}}w^{\rm{J}} p^{\rm{J}}$ is therefore the jamming power cost for each D2D transmitter, where $\lambda_{\cal{D}}A$ is the expected number of D2D transmitters in this region. The inequality constraint (i.e.,~(\ref{eq:D2D_problem_constr})) puts a lower bound on the detection error of warden $\left.w\right|_d$ in order to maintain certain communication covertness for the D2D transmission.

\end{itemize}


\section{Equilibrium Analysis and Algorithm Design of Two-Stage Stackelberg Game}
\label{sec:equilibrium_game}

In this section, we provide the equilibrium analysis for the two-stage Stackelberg game formulated in Section~\ref{subsec:game_formulation} and develop an algorithm to search for the corresponding equilibrium. First, given the strategy from the upper stage, we analyze the lower-stage detection error minimization problem for warden $\left.w\right|_d$ in Section~\ref{subsec:solution_warden} and choose a suitable algorithm to search for the optimal strategy of warden $\left.w\right|_d$ as the best response of the lower stage. Taking into account the best response from the lower stage, we design a bi-level algorithm based on the SCA method to search for the optimal strategy of D2D transmitter $d$ and the jammers at the upper stage, which together with the optimal strategy of warden $\left.w\right|_d$ constitute the equilibrium of the two-stage Stackelberg game. 

\subsection{Solution to Lower-Stage Problem: Detection Error Minimization}
\label{subsec:solution_warden}

\begin{figure}[!]
	\centering
	\includegraphics[width=0.5\textwidth,trim=10 10 5 60, clip]{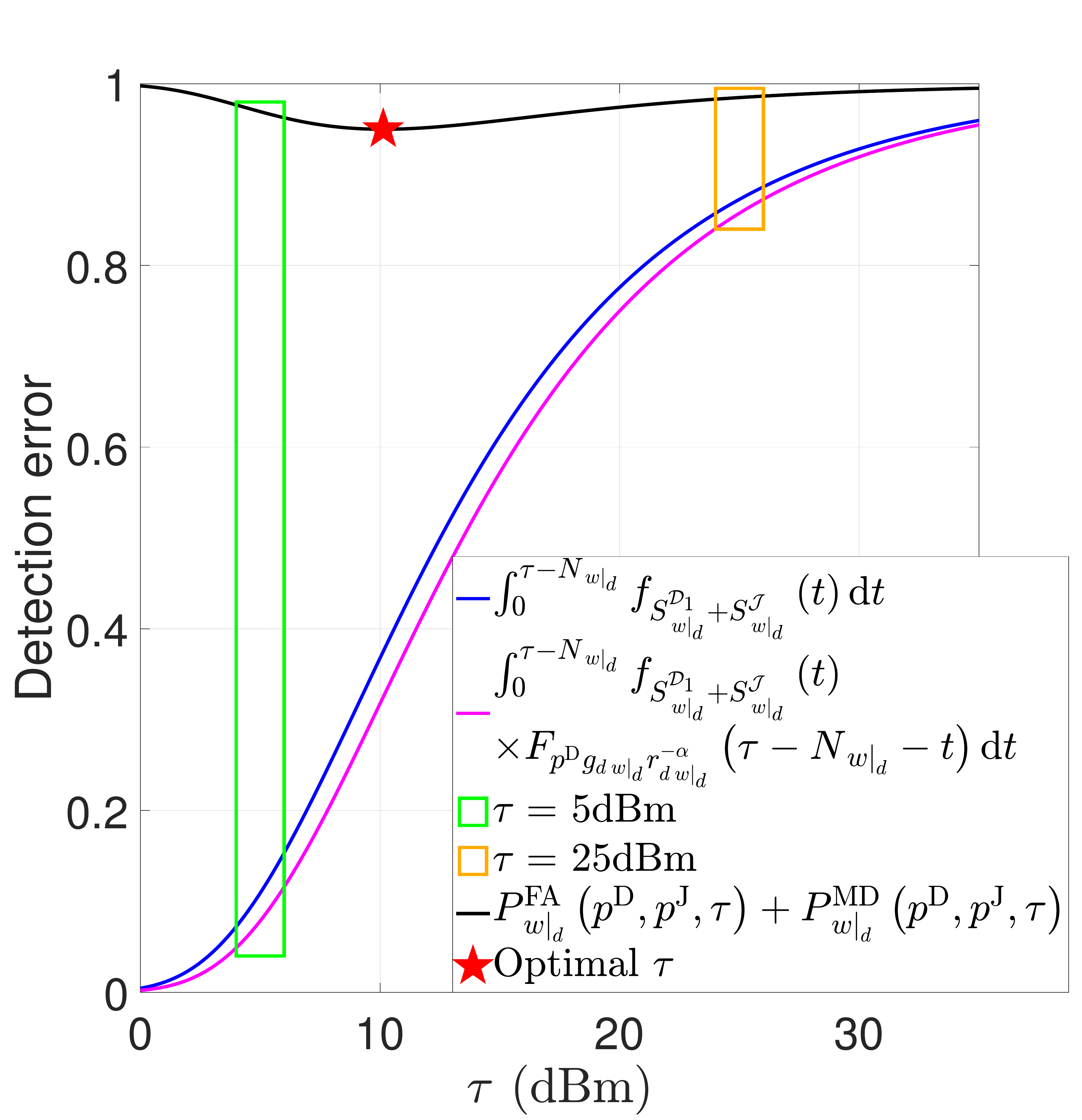}
	\caption{Illustration of the uniqueness of the solution to the lower-stage detection error minimization problem with $p^{\rm{D}} = 20$dBm and $p^{\rm{J}} = 10$dBm.}
	\label{fig:uniqueness_warden}
\end{figure}

For the lower-stage detection error minimization problem in~(\ref{eq:warden_w_problem}), given any D2D transmission power and jamming power (i.e., $p^{\rm{D}}$ and $p^{\rm{J}}$, respectively), the objective of warden $\left.w\right|_d$ will be in a trough shape w.r.t. $\tau$. In this case, we can apply the Rosenbrock method~\cite{leader2004numerical} to obtain the optimal strategy for warden $\left.w\right|_d$. The reason for the trough shape of the warden's objective can be explained as follows. First, we rewrite the objective function of warden $\left.w\right|_d$ (i.e., detection error in~(\ref{eq:warden_w_problem})) as follows:
\begin{equation}\label{eq:warden_w_rewrite}
\begin{aligned}
& {\mathbb{P}}^{\text{FA}}_{\left.w\right|_d} \left(p^{\rm{D}}, p^{\rm{J}}, \tau\right) + {\mathbb{P}}^{\rm{MD}}_{\left.w\right|_d} \left(p^{\rm{D}}, p^{\rm{J}}, \tau\right) \\
= & {\mathbb{P}} \left[S_{\left.w\right|_d}^{{\cal{D}}_1} + S_{\left.w\right|_d}^{{\cal{J}}} + N_{\left.w\right|_d} > \tau \right] + {\mathbb{P}} \left[ p^{\rm{D}} g_{d\left.w\right|_d} r_{d\left.w\right|_d}^{-\alpha} + S_{\left.w\right|_d}^{{\cal{D}}_1} + S_{\left.w\right|_d}^{{\cal{J}}} + N_{\left.w\right|_d} < \tau \right] \\
= & 1 - \int_0^{\tau - N_{\left.w\right|_d}} f_{S_{\left.w\right|_d}^{{\cal{D}}_1} + S_{\left.w\right|_d}^{{\cal{J}}}} \left(t\right) {\rm{d}} t + \int_0^{\tau - N_{\left.w\right|_d}} f_{S_{\left.w\right|_d}^{{\cal{D}}_1} + S_{\left.w\right|_d}^{{\cal{J}}}} \left(t\right) F_{p^{\rm{D}} g_{d\left.w\right|_d} r_{d\left.w\right|_d}^{-\alpha}} \left(\tau - N_{\left.w\right|_d} - t\right) {\rm{d}} t.
\end{aligned}
\end{equation}
In~(\ref{eq:warden_w_rewrite}) and given any D2D transmission power and jamming power (i.e., $p^{\rm{D}}$ and $p^{\rm{J}}$, respectively), as $F_{p^{\rm{D}} g_{d\left.w\right|_d} r_{d\left.w\right|_d}^{-\alpha}} \left(\cdot\right) \in \left[0, 1\right]$, $\int_0^{\tau - N_{\left.w\right|_d}} f_{S_{\left.w\right|_d}^{{\cal{D}}_1} + S_{\left.w\right|_d}^{{\cal{J}}}} \left(t\right) {\rm{d}} t$ will have a steeper increase w.r.t. $\tau$ than $\int_0^{\tau - N_{\left.w\right|_d}} f_{S_{\left.w\right|_d}^{{\cal{D}}_1} + S_{\left.w\right|_d}^{{\cal{J}}}} \left(t\right) F_{p^{\rm{D}} g_{d\left.w\right|_d} r_{d\left.w\right|_d}^{-\alpha}} \left(\tau - N_{\left.w\right|_d} - t\right) {\rm{d}} t$ when $\tau$ is small, which is demonstrated with the numerical results in Fig.~\ref{fig:uniqueness_warden}, e.g., $\tau = 5$dBm. In this case, ${\mathbb{P}}^{\text{FA}}_{\left.w\right|_d} \left(p^{\rm{D}}, p^{\rm{J}}, \tau\right) + {\mathbb{P}}^{\rm{MD}}_{\left.w\right|_d} \left(p^{\rm{D}}, p^{\rm{J}}, \tau\right)$ will decrease w.r.t. $\tau$. Later, when $\tau$ further increases, $\int_0^{\tau - N_{\left.w\right|_d}} f_{S_{\left.w\right|_d}^{{\cal{D}}_1} + S_{\left.w\right|_d}^{{\cal{J}}}} \left(t\right) {\rm{d}} t$ almost approaches its probabilistic upper bound, and its increasing speed becomes slower than that of $\int_0^{\tau - N_{\left.w\right|_d}} f_{S_{\left.w\right|_d}^{{\cal{D}}_1} + S_{\left.w\right|_d}^{{\cal{J}}}} \left(t\right) F_{p^{\rm{D}} g_{d\left.w\right|_d} r_{d\left.w\right|_d}^{-\alpha}} \left(\tau - N_{\left.w\right|_d} - t\right) {\rm{d}} t$, which also can be observed from Fig.~\ref{fig:uniqueness_warden}, e.g., $\tau = 25$dBm. In this case, ${\mathbb{P}}^{\text{FA}}_{\left.w\right|_d} \left(p^{\rm{D}}, p^{\rm{J}}, \tau\right) + {\mathbb{P}}^{\rm{MD}}_{\left.w\right|_d} \left(p^{\rm{D}}, p^{\rm{J}}, \tau\right)$ will increase w.r.t. $\tau$. Consequently, ${\mathbb{P}}^{\text{FA}}_{\left.w\right|_d} \left(p^{\rm{D}}, p^{\rm{J}}, \tau\right) + {\mathbb{P}}^{\rm{MD}}_{\left.w\right|_d} \left(p^{\rm{D}}, p^{\rm{J}}, \tau\right)$ takes on a trough shape w.r.t. $\tau$, which induces a unique optimal strategy (i.e., $\tau^\star$) that minimizes the detection error for warden $\left.w\right|_d$. Such an optimal strategy is the best response of the lower stage corresponding to the given D2D transmission power and jamming power (i.e., $p^{\rm{D}}$ and $p^{\rm{J}}$, respectively).


\subsection{Solution to Upper-Stage Problem: Friendly Jamming-Assisted Covert Communication}
\label{subsec:solution_covert_D2D}

For the friendly jamming-assisted covert communication problem at the upper stage, it is unwieldy to verify its convexity due to the multiple integrals in both the objective and constraint. In this case, we design a bi-level optimization algorithm based on the SCA method~\cite{6675875} and take into account the best response from the lower stage (i.e., $\tau^\star$ in~(\ref{eq:warden_w_problem})). In particular, let $U_0\left(p^{\rm{D}}, p^{\rm{J}}\right)$ be the negative value of~(\ref{eq:D2D_problem_obj}) and $U_1\left(p^{\rm{D}}, p^{\rm{J}}\right) = - {\mathbb{P}}^{\text{FA}}_{\left.w\right|_d} \left(p^{\rm{D}}, p^{\rm{J}}, \tau^\star\right) - {\mathbb{P}}^{\rm{MD}}_{\left.w\right|_d} \left(p^{\rm{D}}, p^{\rm{J}}, \tau^\star\right) + 1 - \varepsilon$, we have the upper-stage friendly jamming-assisted covert communication problem (i.e.,~(\ref{eq:D2D_problem})) rewritten as follows:
\begin{equation}\label{eq:D2D_problem_simplified}
\begin{aligned}
\min_{p^{\rm{D}}, p^{\rm{J}}} & \, U_0\left(p^{\rm{D}}, p^{\rm{J}}\right)\\
{\text{s.t.}} & \, U_1\left(p^{\rm{D}}, p^{\rm{J}}\right) \le 0 \\
& \, p^{\rm{D}} \in \left[{\underline{p}}^{\rm{D}}, {\overline{p}}^{\rm{D}}\right], \, p^{\rm{J}} \in \left[{\underline{p}}^{\rm{J}}, {\overline{p}}^{\rm{J}}\right].
\end{aligned}
\end{equation}
According to~\cite{6675875}, we can follow Algorithm~\ref{alg:bi_level} and switch to solve an approximation convex program of~(\ref{eq:D2D_problem_simplified}) (i.e.,~(\ref{eq:D2D_problem_approximation})) iteratively. The convergence of Algorithm~\ref{alg:bi_level} has been proven in~\cite{Daniel2020algorithms}.~(\ref{eq:D2D_problem_approximation}) is formulated by replacing $U_i\left(p^{\rm{D}}, p^{\rm{J}}\right)$ for all $i \in \left\{0,\,1\right\}$ of~(\ref{eq:D2D_problem_simplified}) with the surrogate function ${\hat{U}}_i\left(\left.{p^{\rm{D}}}^{\langle k \rangle}, {p^{\rm{J}}}^{\langle k \rangle}\right| {p^{\rm{D}}}^{\langle k \rangle}, {p^{\rm{J}}}^{\langle k \rangle}\right)$ for all $i\in\left\{0,\,1\right\}$ that satisfies the following conditions~\cite{Daniel2020algorithms}:
\begin{enumerate}
\item Consistency in function value: ${\hat{U}}_i\left(\left.{p^{\rm{D}}}^{\langle k \rangle}, {p^{\rm{J}}}^{\langle k \rangle}\right| {p^{\rm{D}}}^{\langle k \rangle}, {p^{\rm{J}}}^{\langle k \rangle}\right) = U_i\left({p^{\rm{D}}}^{\langle k \rangle}, {p^{\rm{J}}}^{\langle k \rangle}\right)$;

\item Consistency in gradient: $\nabla_{p^{\rm{D}}} {\hat{U}}_i\left(\left.{p^{\rm{D}}}^{\langle k \rangle}, {p^{\rm{J}}}^{\langle k \rangle}\right| {p^{\rm{D}}}^{\langle k \rangle}, {p^{\rm{J}}}^{\langle k \rangle}\right) = \nabla_{p^{\rm{D}}} U_i\left({p^{\rm{D}}}^{\langle k \rangle}, {p^{\rm{J}}}^{\langle k \rangle}\right)$ and $\nabla_{p^{\rm{J}}} {\hat{U}}_i\left(\left.{p^{\rm{D}}}^{\langle k \rangle}, {p^{\rm{J}}}^{\langle k \rangle}\right| {p^{\rm{D}}}^{\langle k \rangle}, {p^{\rm{J}}}^{\langle k \rangle}\right) = \nabla_{p^{\rm{J}}} U_i\left({p^{\rm{D}}}^{\langle k \rangle}, {p^{\rm{J}}}^{\langle k \rangle}\right)$;

\item Convexity: ${\hat{U}}_i\left(\left.p^{\rm{D}}, p^{\rm{J}}\right| {p^{\rm{D}}}^{\langle k \rangle}, {p^{\rm{J}}}^{\langle k \rangle}\right)$ is strongly convex w.r.t. $\left[p^{\rm{D}}, p^{\rm{J}}\right]$.
\end{enumerate}
As stated in~\cite{Daniel2020algorithms}, the surrogate function ${\hat{U}}_i\left(\left.p^{\rm{D}}, p^{\rm{J}}\right| {p^{\rm{D}}}^{\langle k \rangle}, {p^{\rm{J}}}^{\langle k \rangle}\right)$ for all $i\in\left\{0, 1\right\}$ in~(\ref{eq:D2D_problem_approximation}) can be designed by using the following gradient descent update:
\begin{equation}\label{eq:surrogate_function}
\begin{aligned}
{\hat{U}}_i\left(\left.p^{\rm{D}}, p^{\rm{J}}\right| {p^{\rm{D}}}^{\langle k \rangle}, {p^{\rm{J}}}^{\langle k \rangle}\right) = & U_i\left({p^{\rm{D}}}^{\langle k \rangle}, {p^{\rm{J}}}^{\langle k \rangle}\right) + \nabla U_i\left({p^{\rm{D}}}^{\langle k \rangle}, {p^{\rm{J}}}^{\langle k \rangle}\right)^\top  \begin{pmatrix}
   p^{\rm{D}} - {p^{\rm{D}}}^{\langle k \rangle} \\
   p^{\rm{J}} - {p^{\rm{J}}}^{\langle k \rangle} \\
\end{pmatrix} \\
& + \frac{\delta}{2}\left(p^{\rm{D}} - {p^{\rm{D}}}^{\langle k \rangle}\right)^2  + \frac{\delta}{2}\left(p^{\rm{J}} - {p^{\rm{J}}}^{\langle k \rangle}\right)^2.
\end{aligned}
\end{equation}

\begin{algorithm}[H]
 \KwData{Input ${p^{\rm{D}}}^{\langle k \rangle}$ and ${p^{\rm{J}}}^{\langle k \rangle}$.}
 Initialize $k = 0$ and set error tolerance $\epsilon$\;
 \While{error $ > \epsilon$}{
Solve the following optimization problem
  \begin{equation}\label{eq:D2D_problem_approximation}
	\begin{aligned}
	\min_{p^{\rm{D}}, p^{\rm{J}}} & \, {\hat{U}}_0\left(\left.p^{\rm{D}}, p^{\rm{J}}\right|{p^{\rm{D}}}^{\langle k \rangle}, {p^{\rm{J}}}^{\langle k \rangle}\right)\\
	{\text{s.t.}} & \, {\hat{U}}_1\left(\left.p^{\rm{D}}, p^{\rm{J}}\right|{p^{\rm{D}}}^{\langle k \rangle}, {p^{\rm{J}}}^{\langle k \rangle}\right) \le 0 \\
	& \, p^{\rm{D}} \in \left[{\underline{p}}^{\rm{D}}, {\overline{p}}^{\rm{D}}\right], \, p^{\rm{J}} \in \left[{\underline{p}}^{\rm{J}}, {\overline{p}}^{\rm{J}}\right]
	\end{aligned}
  \end{equation}
  and obtain the solution ${\tilde{p}}^{\rm{D}}$ and ${\tilde{p}}^{\rm{J}}$\;
  ${p^{\rm{D}}}^{\langle k + 1 \rangle} = {p^{\rm{D}}}^{\langle k \rangle} + \gamma \left({\tilde{p}}^{\rm{D}} - {p^{\rm{D}}}^{\langle k \rangle}\right)$ and ${p^{\rm{J}}}^{\langle k + 1 \rangle} = {p^{\rm{J}}}^{\langle k \rangle} + \gamma \left({\tilde{p}}^{\rm{J}} - {p^{\rm{J}}}^{\langle k \rangle}\right)$ \;
  $error = \left\|{p^{\rm{D}}}^{\langle k + 1 \rangle} - {p^{\rm{D}}}^{\langle k \rangle}\right\|^2 + \left\|{p^{\rm{J}}}^{\langle k + 1 \rangle} - {p^{\rm{J}}}^{\langle k \rangle}\right\|^2$\;
  $k + 1 \mapsto k$\;
 }
 \caption{Bi-level Optimization Algorithm}
 \label{alg:bi_level}
\end{algorithm}


\section{Performance Evaluation}
\label{sec:performance}

In this section, we provide the numerical results to evaluate the system performance. The parameter setting is given in Table~\ref{tab:notation_value}. First, we verify the optimality of the obtained strategy and the effectiveness of the friendly jamming in Fig.~\ref{fig:D2D_utility_algo_convergence}. Second, we demonstrate the advantages of the proposed friendly jamming-assisted covert communication by comparing it to the information-theoretical secrecy approach in Fig.~\ref{fig:covert_vs_secrecy}. Third, the impact of the communication covertness requirement and the warden's density on the network utility are illustrated in Fig.~\ref{fig:impact_varepsilon_lambda_W}. Fourth, we evaluate the impact of the densities of the D2D transmitters and jammers on the network utility in Fig.~\ref{fig:impact_lambda_D_J}. Finally, we present the impact of the SINR thresholds at the warden and the D2D transmitter on the network utility in Fig.~\ref{fig:impact_xi_D_W}.

\subsection{Effectiveness of Friendly Jamming}

\begin{figure}
     \centering
     \begin{minipage}{8cm}
		\centering
		\includegraphics[width=1\textwidth,trim=10 0 10 20,clip]{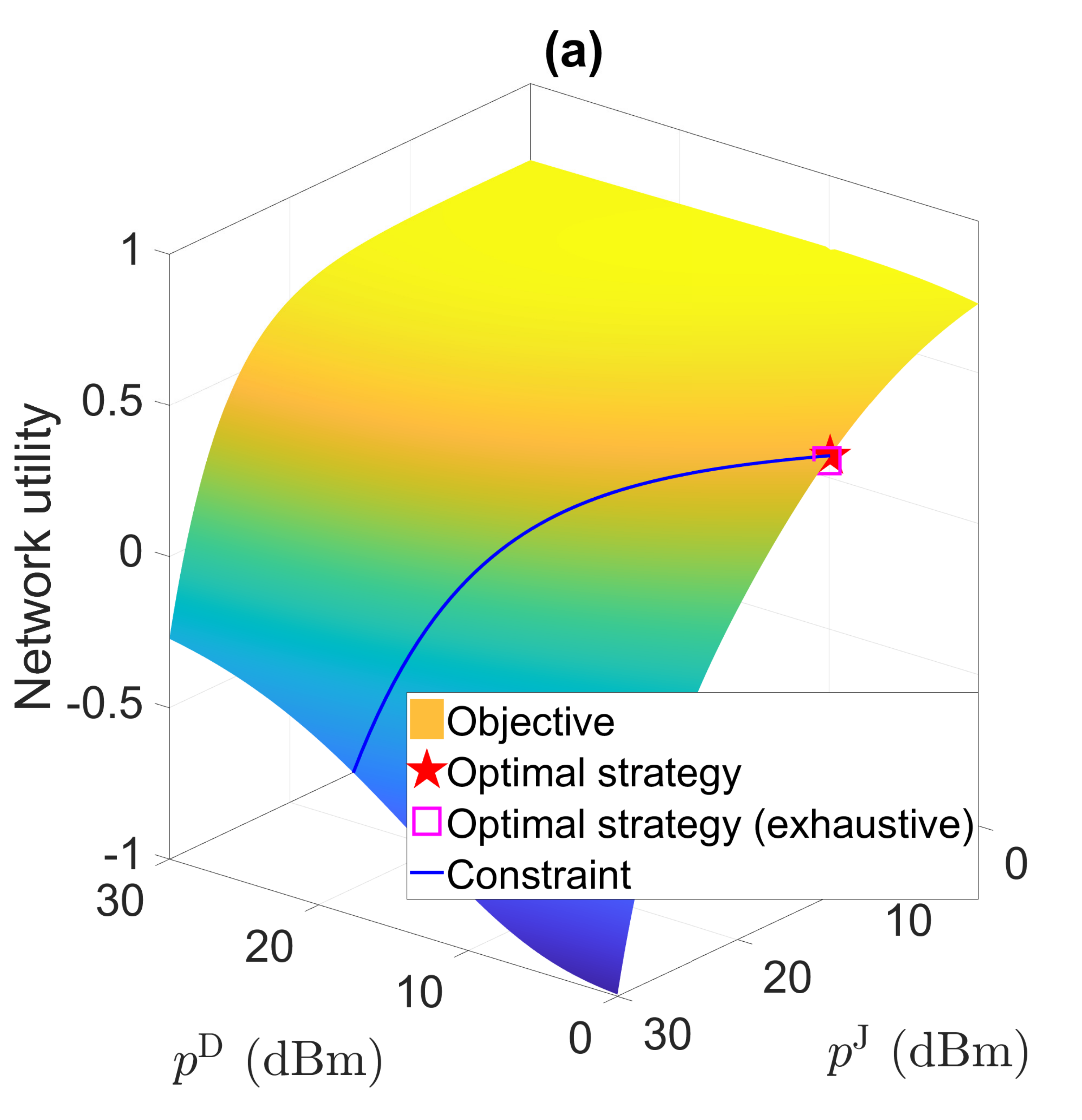}
     \end{minipage}
     \begin{minipage}{8cm}
		\centering
		\includegraphics[width=1\textwidth,trim=10 0 10 20,clip]{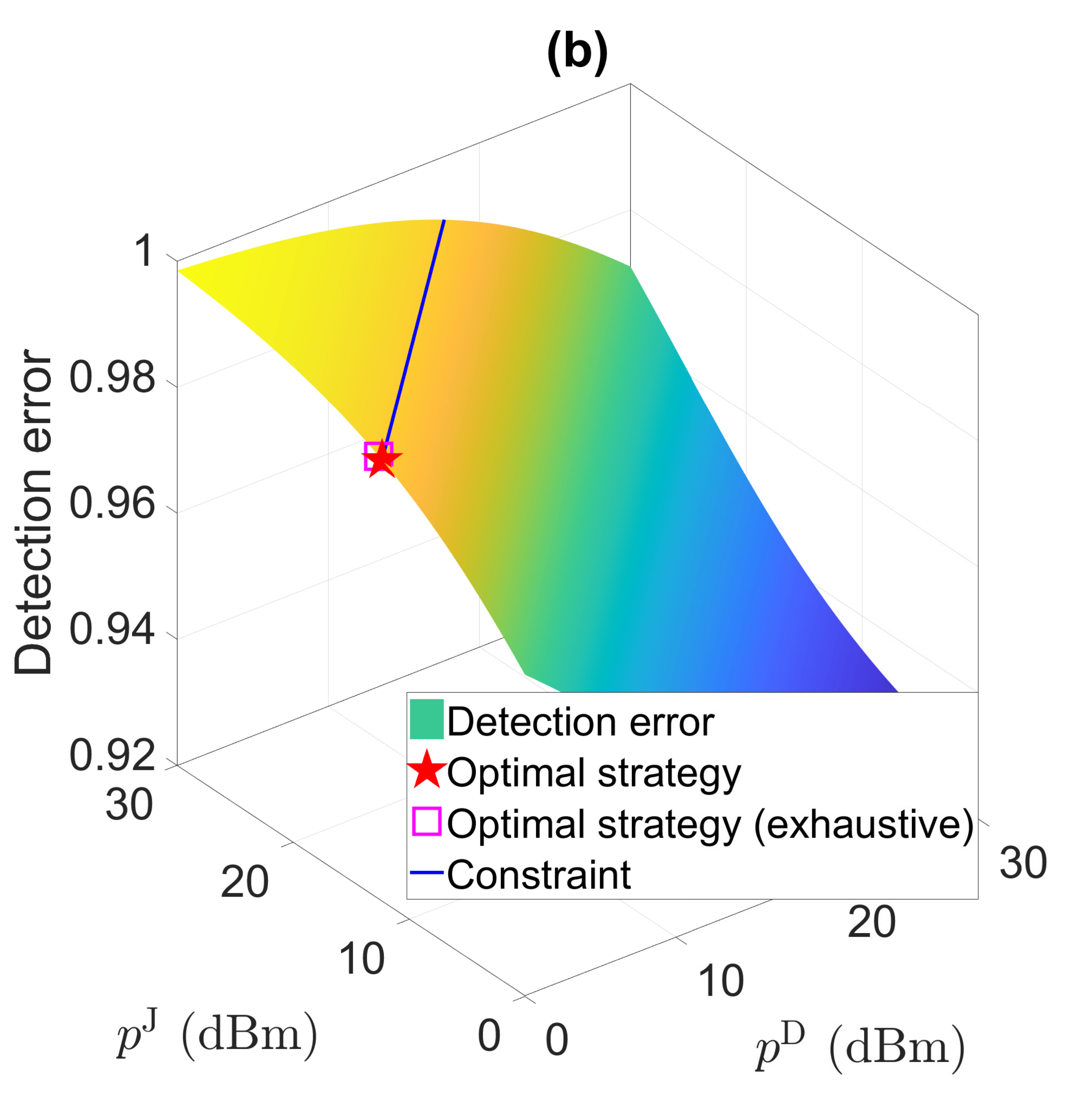}
     \end{minipage}
        \caption{(a) Network utility and (b) detection error w.r.t. D2D transmission power and jamming power (i.e., $p^{\rm{D}}$ and $p^{\rm{J}}$, respectively).}
        \label{fig:D2D_utility_algo_convergence}
\end{figure}

Figure~\ref{fig:D2D_utility_algo_convergence} depicts the network utility and detection error as functions of the D2D transmission power $p^{\rm{D}}$ and jamming power $p^{\rm{J}}$. In Fig.~\ref{fig:D2D_utility_algo_convergence}(a), we can observe that the optimal strategy (i.e., optimal D2D transmission power and jamming power) can maximize the network utility within the feasible domain, where the boundary of the feasible domain is determined by the constraint on communication covertness (i.e.,~(\ref{eq:D2D_problem_constr})) as shown in Fig.~\ref{fig:D2D_utility_algo_convergence}(b). The optimal strategy is obtained by the developed bi-level optimization algorithm (i.e., Algorithm~\ref{alg:bi_level}) and is consistent with that obtained by the exhaustive search method. The exhaustive search method searches for the optimal strategy via: 1) discretizing the domain of definition of the strategy (i.e., the D2D transmission power and jamming power), 2) enumerating all possible candidates for the strategy taking into account the corresponding best response from the lower stage to evaluate the network utility and detection error, and 3) finding the optimal strategy that jointly maximizes the network utility and satisfies the constraint on communication covertness. In this case, the optimality of the obtained equilibrium strategy is validated. Moreover, as shown in Fig.~\ref{fig:D2D_utility_algo_convergence}(b), the detection error can satisfy the constraint on communication covertness (i.e.,~(\ref{eq:D2D_problem_constr})) if and only if the jamming power $p^{\rm{J}} \gtrsim 12$dBm\footnote{$p^{\rm{J}}$ is approximately larger than $12$dBm.}. The reason is that merely relying on the co-channel interference in the D2D network is insufficient and hence difficult to confuse and mislead the decision of the warden. In this case, without adequate assistance from the friendly jammers, the covert communication cannot be achieved.



\subsection{Advantages of Friendly Jamming-Assisted Covert Communication}

\begin{figure}
		\centering
		\includegraphics[width=0.5\textwidth,trim=10 0 0 10,clip]{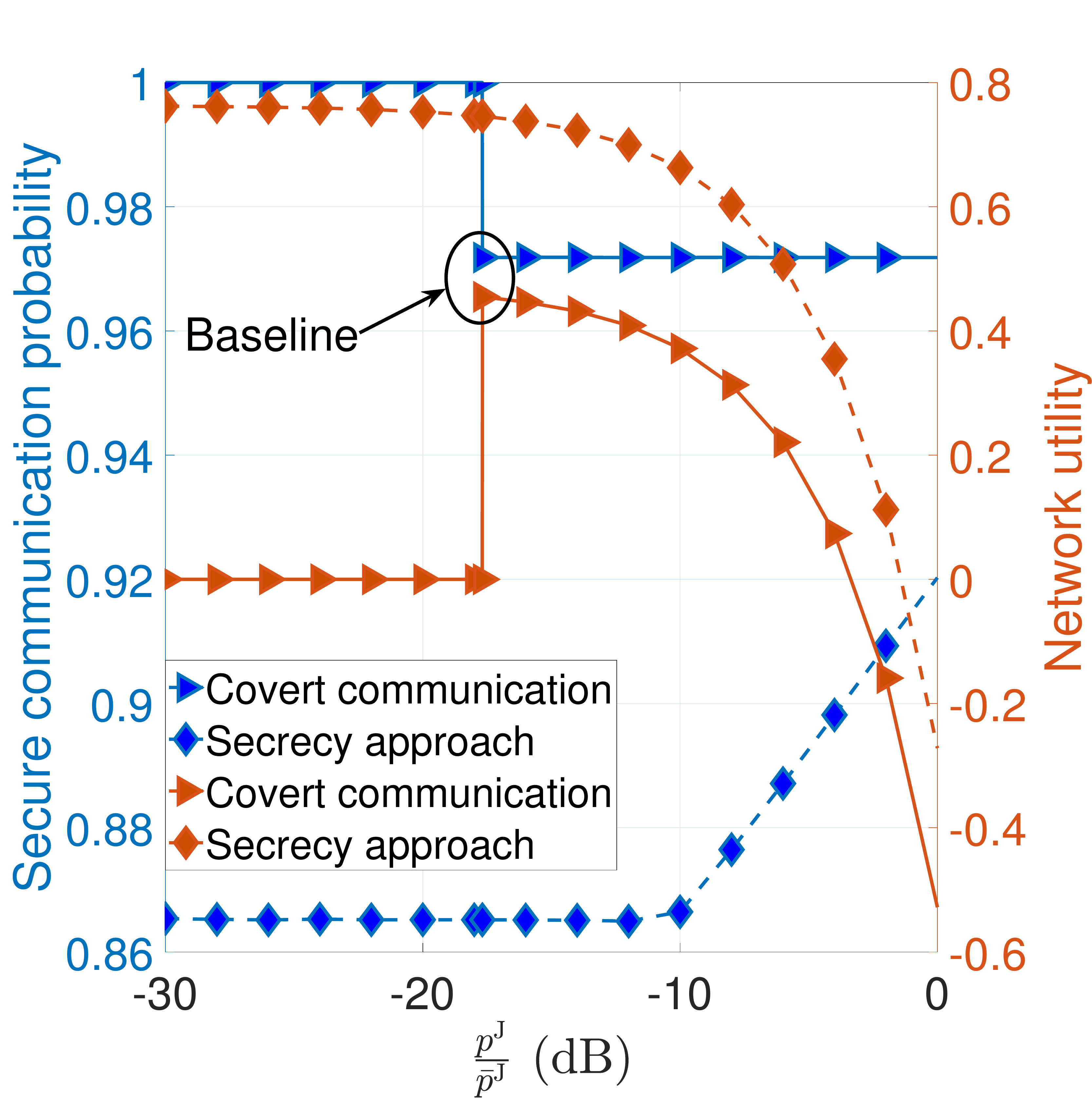}
        \caption{Friendly jamming-assisted covert communication versus information-theoretical secrecy approach.}
        \label{fig:covert_vs_secrecy}
\end{figure}

In Fig.~\ref{fig:covert_vs_secrecy}, we compare our proposed friendly jamming-assisted covert communication and the information-theoretic secrecy approach in terms of the network utility and the secure communication probability (i.e.,~(\ref{eq:prob_secure_communication})). The problem of the information-theoretic secrecy approach is the problem in~(\ref{eq:D2D_problem}) with the removal of the constraint on communication covertness (i.e.,~(\ref{eq:D2D_problem_constr})). By such, the problem of the information-theoretic secrecy approach aims at the joint link reliability maximization and secrecy outage minimization. We can observe that the friendly jamming-assisted covert communication starts to take effect at $\frac{p^{\rm{J}}}{{\bar{p}}^{\rm{J}}} \approx -18$dB, where the network utility becomes nonzero and also the optimal performance is achieved (i.e., the network utility is maximized at around $0.5$ with a certain secure communication probability at around $0.97$). In this case, the optimal performance of the friendly jamming-assisted covert communication at $\frac{p^{\rm{J}}}{{\bar{p}}^{\rm{J}}} \approx -18$dB will be used as a ``{\bf{Baseline}}'' for comparison in the following discussion. Regarding the information-theoretical secrecy approach, initially it can achieve a higher utility and a lower secure communication probability compared to the optimal performance of the friendly jamming-assisted covert communication. Later, if we further increase the jamming power, although the secure communication probability of the information-theoretical secrecy approach significantly increases, its network utility decreases dramatically. Until the jamming power approaches the upper bound, the secure communication probability of the information-theoretical secrecy approach is still lower than that of the friendly jamming-assisted covert communication at $\frac{p^{\rm{J}}}{{\bar{p}}^{\rm{J}}} \approx -18$dB (i.e., around $0.97$) while achieving a negative utility, which is much lower than the utility of the friendly jamming-assisted covert communication at $\frac{p^{\rm{J}}}{{\bar{p}}^{\rm{J}}} \approx -18$dB (i.e., around $0.5$). In this case, the friendly jamming-assisted covert communication significantly outperforms the information-theoretical secrecy approach, and its advantages are well demonstrated.


\subsection{Impact of Warden's Density and Communication Covertness Requirement}

\begin{figure}
     \centering
		\includegraphics[width=0.5\textwidth,trim=0 0 30 10,clip]{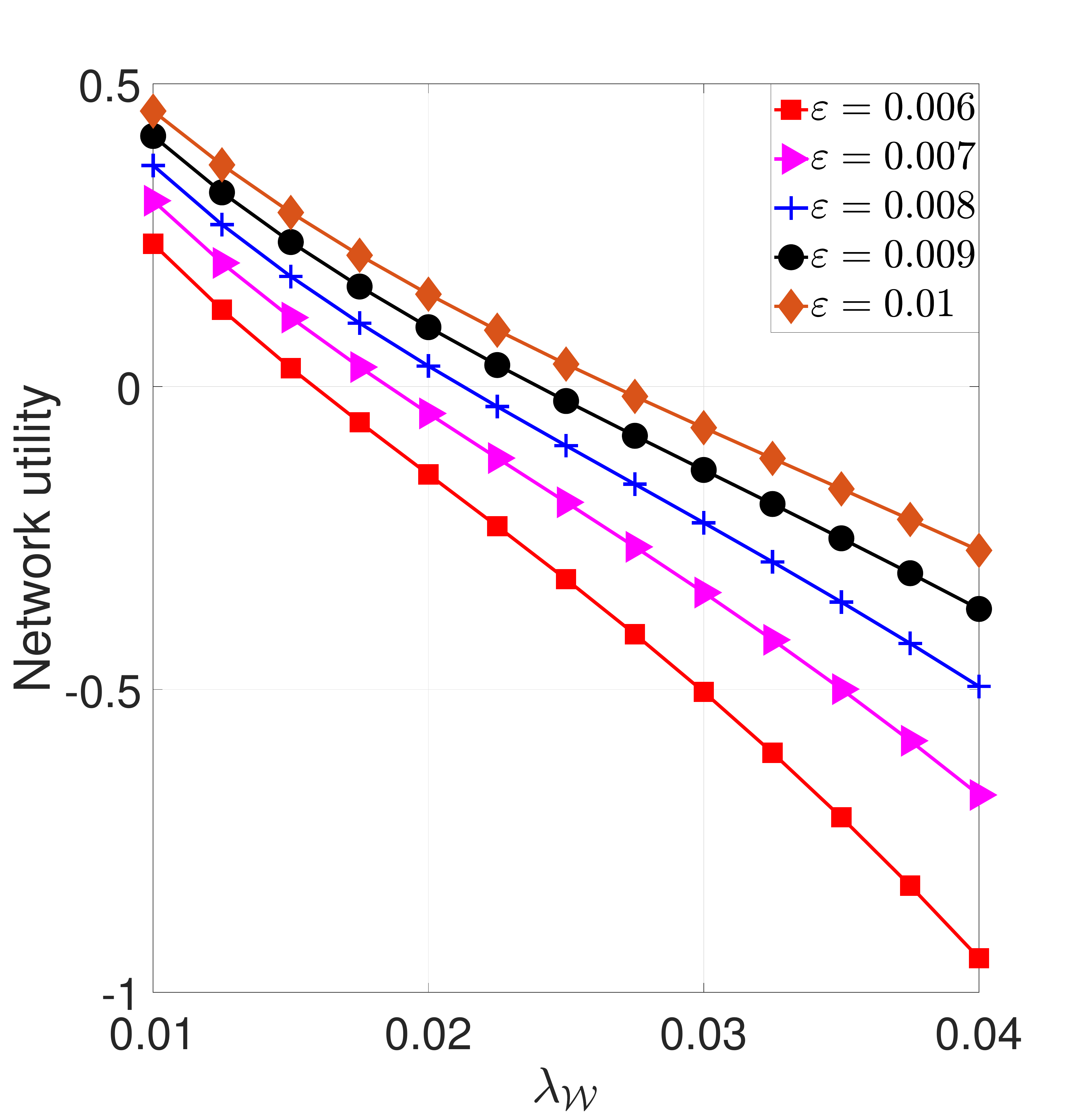}
        \caption{Network utility w.r.t. the warden's density $\lambda_{{\cal{W}}}$ with varying communication covertness requirement $\varepsilon$.}
        \label{fig:impact_varepsilon_lambda_W}
\end{figure}

We evaluate the impact of the warden's density $\lambda_{{\cal{W}}}$ and the communication covertness requirement $\varepsilon$, as given in the nonlinear inequality constraint~(\ref{eq:D2D_problem_constr}), on the network utility in Fig.~\ref{fig:impact_varepsilon_lambda_W}. The network utility decreases w.r.t. the increase in the warden's density $\lambda_{{\cal{W}}}$. The reason is that when the warden's density becomes higher, the distance between the warden and its target D2D transmitter becomes shorter. In this case, the large-scale fading caused by the path loss weakens, the signal power received at the warden from its target D2D transmitter becomes stronger, which simultaneously reduces the difficulty in the transmission detection and increases the secrecy outage probability for D2D transmission once being detected. As a result, the network utility deteriorates accordingly. Additionally, the network utility decreases w.r.t. the warden's density $\lambda_{{\cal{W}}}$ with different rates corresponding to different communication covertness requirements (i.e., $\varepsilon$). This is due to the fact that when the value of $\varepsilon$ decreases, the communication covertness requirement is tighter and it becomes more difficult to achieve covert communication, which together with the more threatening warden (i.e., higher warden densification) induce a faster decreasing rate in the network utility. 


\subsection{Impact of Densities of Friendly Jammers and D2D Transmitters}

\begin{figure}
     \centering
		\includegraphics[width=0.5\textwidth,trim=0 0 30 10,clip]{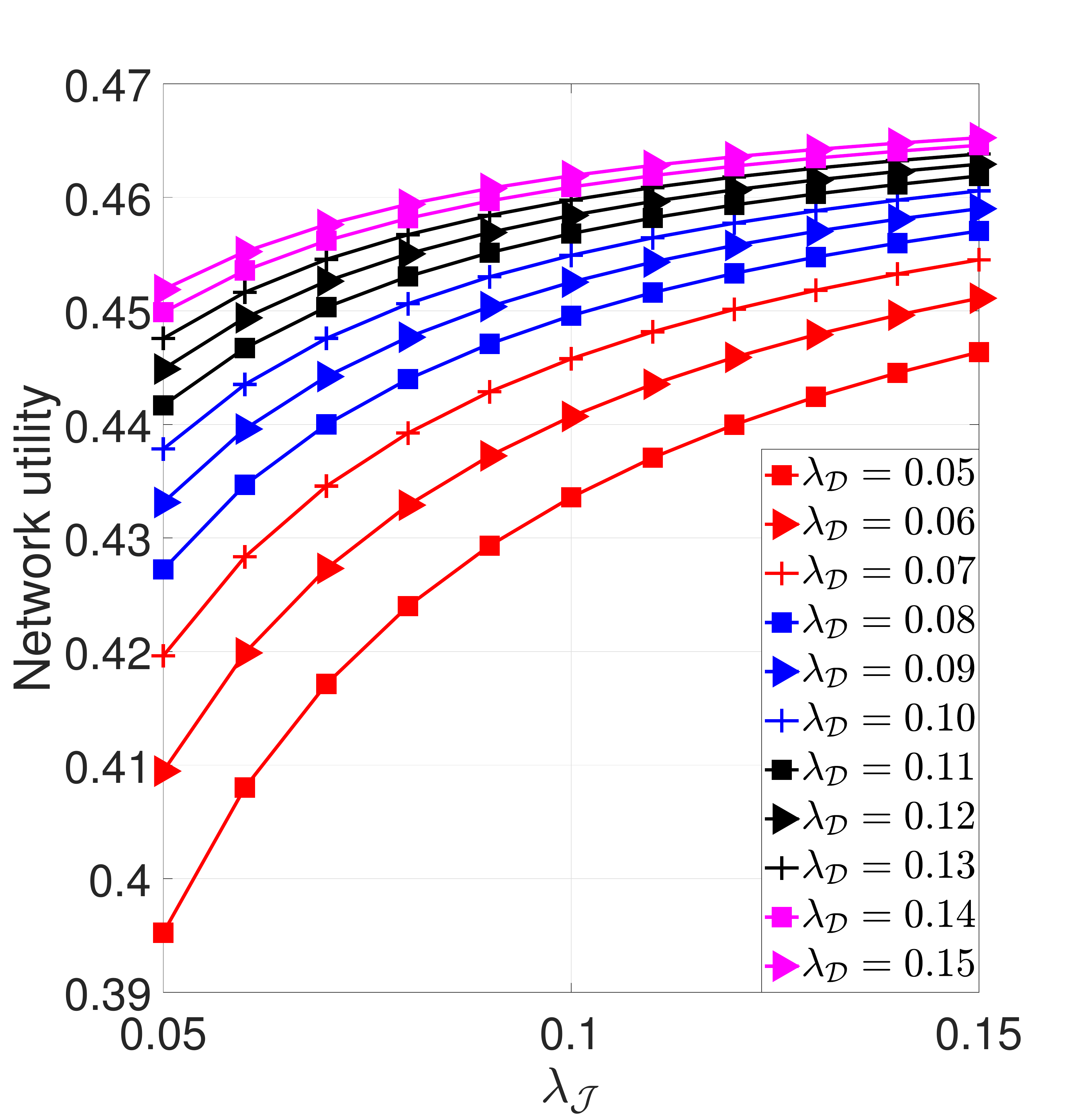}
        \caption{Network utility w.r.t. the jammers's density $\lambda_{{\cal{J}}}$ with varying D2D transmitters' density $\lambda_{\cal{D}}$.}
        \label{fig:impact_lambda_D_J}
\end{figure}

Figure~\ref{fig:impact_lambda_D_J} evaluates the impact of the densities of the friendly jammers and D2D transmitters on the network utility. An interesting result can be observed from Fig.~\ref{fig:impact_lambda_D_J} that the increase in the jammers' density improves instead of damaging the network utility. The reason is that although the increase in the jammers' density leads to stronger interference to the D2D communication and weakens the SINR at the D2D receiver, this interference can also enhance the communication covertness. Based on the results shown in Fig.~\ref{fig:impact_lambda_D_J}, the gain obtained from the enhancement of communication covertness overcompensates the loss incurred by the weakened SINR. This again demonstrates the benefit of using the friendly jammers in achieving covert communication. Another interesting result can be observed in Fig.~\ref{fig:impact_lambda_D_J} that the improvement due to the increase in jammers' density becomes diminishing when the D2D transmitters' density increases. The reason is that as the density of D2D transmitters increases, the co-channel interference due to the concurrent transmission becomes stronger, which can also distort the observation of the warden and mislead its decision. In this case, it is not necessary to have massive jammers as the assistants in achieving covert communication. Hence, the improvement due to the increase in jammers' density becomes less when the D2D transmitters' density increases.


\subsection{Impact of SINR Thresholds at Warden and D2D Transmitter}

\begin{figure}
     \centering
		\includegraphics[width=0.5\textwidth,trim=0 0 30 10,clip]{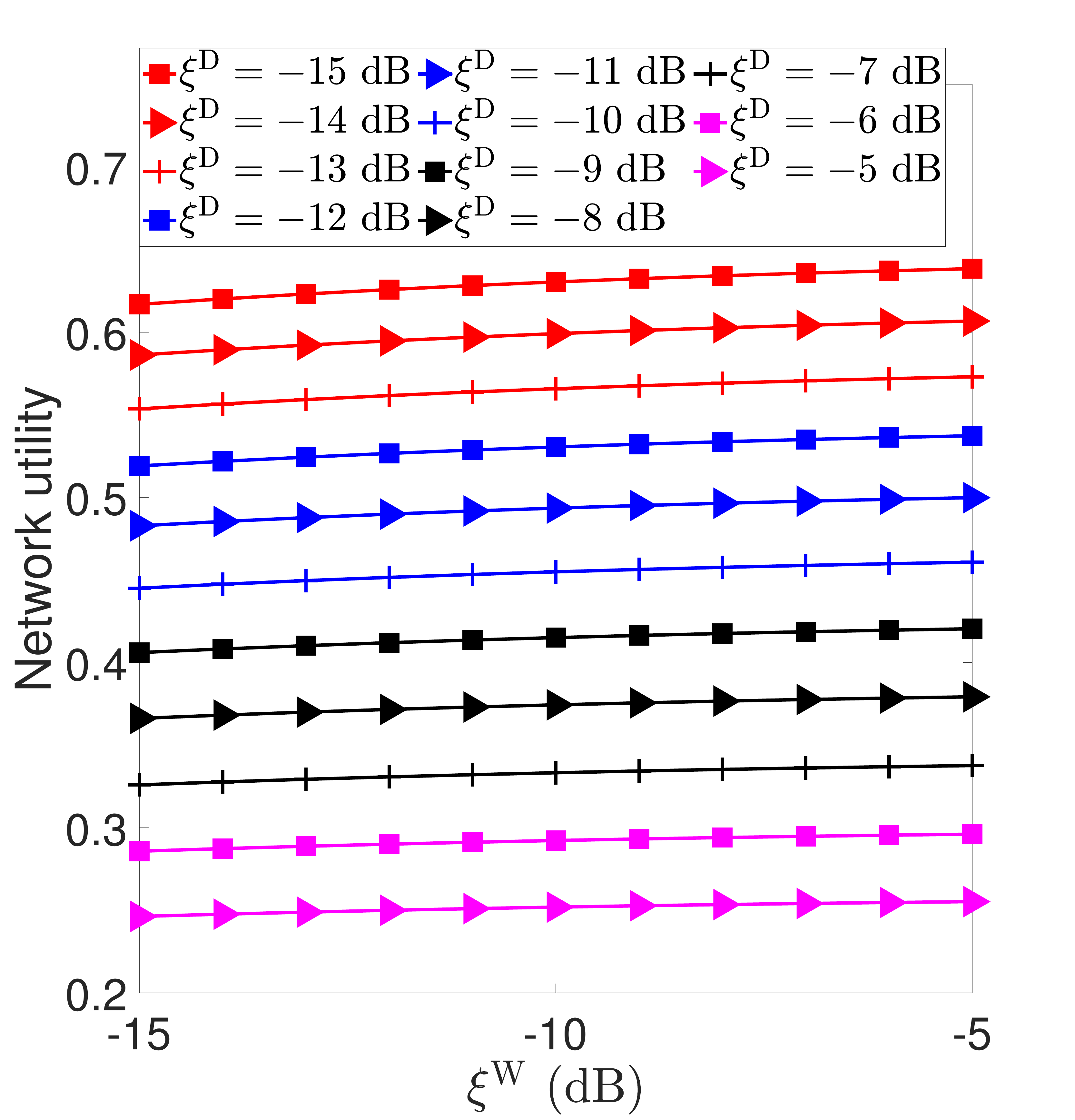}
        \caption{Network utility w.r.t. the SINR threshold at warden $\xi^{{\rm{W}}}$ with varying SINR threshold at D2D receiver $\xi^{\rm{D}}$.}
        \label{fig:impact_xi_D_W}
\end{figure}

We evaluate the impact of the SINR threshold at warden and D2D transmitter (i.e., $\xi^{{\rm{W}}}$ and $\xi^{\rm{D}}$, respectively). As shown in Fig.~\ref{fig:impact_xi_D_W}, the network utility concavely increases w.r.t. $\xi^{{\rm{W}}}$. This is due to the fact that to successfully decode the received signal is of more difficulty for the warden along with the increase in the SINR threshold $\xi^{{\rm{W}}}$. In this case, the secure communication probability (i.e.,~(\ref{eq:prob_secure_communication})) will improve, which induces a higher network utility. The concave increasing trend in the network utility regarding the increase in $\xi^{{\rm{W}}}$ can be explained as follows. With the increase in $\xi^{{\rm{W}}}$, the secrecy outage probability gradually approaches its probabilistic lower bound, and its decreasing speed becomes slower than before. In this case, the increasing speed in the secure communication probability becomes slower, which induces a concave increasing trend in the network utility. Regarding the improvement in the network utility corresponding to the decrease in the SINR threshold at D2D transmitter $\xi^{{\rm{D}}}$. This is due to that the decrease in $\xi^{{\rm{D}}}$ can improve the successful transmission probability for the D2D transmitter (i.e.,~(\ref{eq:ProbRel_SINR})), which makes the objective of the D2D network easier in achieving higher value. 


\section{Conclusion}
\label{sec:conclusion}

We leverage both covert communication and friendly jamming to achieve physical-layer security in a large-scale D2D network in order to defend against the wardens.  The objective is to jointly increase the detection error at the warden and decrease secrecy outage for the D2D transmitters in case transmissions are detected. The  combat between the wardens and the D2D network (i.e., the D2D transmitters and the jammers) is modeled by a Stackelberg game with the wardens as the followers at the lower stage and the legitimate entities as the leaders at the upper stage. For the problem of the warden at the lower stage, we both analytically and numerically validate the existence and uniqueness of its optimal strategy, which is regarded as the best response from the lower stage. Given the best response from the lower stage, we develop a bi-level optimization algorithm based on SCA method to search for the optimal strategy of the D2D network, which together with that of the warden constitute the Stackelberg equilibrium. We  have presented numerical results to evaluate the proposed approach and validate the optimality of the obtained equilibrium strategy. Moreover, we have demonstrated the advantages of the proposed friendly jamming-assisted covert communication by comparing it with the information-theoretic secrecy approach in terms of network utility and secure communication probability. Furthermore, we have evaluated the impact of warden's density and communication covertness requirement on system performance. Considering multiple antennas at both the D2D transmitters and receivers and also using the proposed approach to secure D2D communication in a D2D-underlaid cellular network will be future extensions of this work.


\begin{appendices}

\section{SINR Distribution for D2D Network}
\label{app:SINR_d2d}

Based on~(\ref{eq:ProbRel_SINR}) and conditioned on ${\cal{D}}_1$, the successful transmission probability for D2D transmitter $d$ is  
\begin{equation}\label{eq:Prob_SINR_specific}
\begin{aligned}
& {\mathbb{P}} \left[\left.{\rm{SINR}}_d \left(p^{\rm{D}}, p^{\rm{J}}\right) \ge \xi^{\rm{D}} \right|{\cal{D}}_1\right] = {\mathbb{P}} \left[  \frac{p^{\rm{D}} g_{d} R^{-\alpha}}{ S_d^{{\cal{D}}_1} + S_d^{{\cal{J}}} + N_d}\ge \xi^{\rm{D}} \right]\\
= & {\mathbb{P}} \left[ g_d \ge \frac{R^{\alpha} \xi^{\rm{D}}} {p^{\rm{D}}} \left(S_d^{{\cal{D}}_1} + S_d^{{\cal{J}}} + N_d\right) \right]\\ 
\mathop = \limits^{(a)} & {\mathbb{E}} \left[\exp \left( - \frac{R^{\alpha} \xi^{\rm{D}}} {p^{\rm{D}}} \left(S_d^{{\cal{D}}_1} + S_d^{{\cal{J}}} + N_d\right)\right) \right]\\ 
\mathop = \limits^{(b)} & \exp \left( - \frac{R^{\alpha} \xi^{\rm{D}}} {p^{\rm{D}}} N_d\right) {\mathbb{E}}_{S_d^{{\cal{D}}_1}} \left[\exp \left( - \frac{R^{\alpha} \xi^{\rm{D}}} {p^{\rm{D}}} S_d^{{\cal{D}}_1}\right) \right] {\mathbb{E}}_{S_d^{{\cal{J}}}} \left[\exp \left( - \frac{R^{\alpha} \xi^{\rm{D}}} {p^{\rm{D}}} S_d^{{\cal{J}}}\right) \right],
\end{aligned}
\end{equation}
where $(a)$ follows from the Rayleigh fading of $g_d$ and $(b)$ follows the independence between the random variables $S_d^{{\cal{D}}_1}$ and $S_d^{{\cal{J}}}$. The second term of the final row of~(\ref{eq:Prob_SINR_specific}) (i.e., ${\mathbb{E}}_{S_d^{{\cal{D}}_1}} \left[\exp \left( - \frac{R^{\alpha} \xi^{\rm{D}}} {p^{\rm{D}}} S_d^{{\cal{D}}_1}\right) \right]$) is derived as follows:
\begin{equation}\label{eq:Prob_SINR_specific_second_term}
\begin{aligned}
& {\mathbb{E}}_{S_d^{{\cal{D}}_1}} \left[\exp \left( - \frac{R^{\alpha} \xi^{\rm{D}}} {p^{\rm{D}}} S_d^{{\cal{D}}_1}\right) \right] = {\mathbb{E}} \left[\exp \left( - \frac{R^{\alpha} \xi^{\rm{D}}} {p^{\rm{D}}} p^{\rm{D}} \sum_{d' \in \Phi_{\left\{\left.{\cal{D}} \backslash \left\{d\right\}\right| d\right\}}} {\mathbbm{1}}_{d'} g_{d'd} r_{d'd}^{-\alpha}\right) \right]\\
\mathop = \limits^{(a)} & {\mathbb{E}}_{\Phi_{\left\{\left.{\cal{D}} \backslash \left\{d\right\}\right| d\right\}}} \left[\prod_{d' \in \Phi_{\left\{\left.{\cal{D}} \backslash \left\{d\right\}\right| d\right\}}} \left(\frac{{\mathbb{P}}^{{\cal{D}}_1}}{R^{\alpha} \xi^{\rm{D}} r_{d'd}^{-\alpha} + 1} + 1 - {\mathbb{P}}^{{\cal{D}}_1} \right) \right]\\
\mathop = \limits^{(b)} & \exp\left( -2\pi \lambda_{\cal{D}} \int^\infty_0 \left[1 - \left(\frac{{\mathbb{P}}^{{\cal{D}}_1}}{R^{\alpha} \xi^{\rm{D}} r_{d'd}^{-\alpha} + 1} + 1 - {\mathbb{P}}^{{\cal{D}}_1} \right) \right]r_{d'd} {\rm{d}} r_{d'd}\right)\\
= & \exp\left( -2\pi \lambda_{\cal{D}} {\mathbb{P}}^{{\cal{D}}_1} R^2 \left(\xi^{\rm{D}}\right)^{\frac{2}{\alpha}} \int^\infty_0 \frac{r}{1 + r^\alpha} {\rm{d}} r \right)\\
= & \exp\left( -2\pi \lambda_{\cal{D}} {\mathbb{P}}^{{\cal{D}}_1} R^2 \left(\xi^{\rm{D}}\right)^{\frac{2}{\alpha}} \int^\infty_0 r \int^\infty_0 \exp\left(-t\left(1 + r^\alpha\right)\right) {\rm{d}} t {\rm{d}} r \right)\\
= & \exp\left( -2\pi \lambda_{\cal{D}} {\mathbb{P}}^{{\cal{D}}_1} R^2 \left(\xi^{\rm{D}}\right)^{\frac{2}{\alpha}} \int^\infty_0 \exp\left(-t\right) \int^\infty_0 r\exp\left(-tr^\alpha\right) {\rm{d}} r {\rm{d}} t \right)\\
= & \exp\left( -2\pi \lambda_{\cal{D}} {\mathbb{P}}^{{\cal{D}}_1} R^2 \left(\xi^{\rm{D}}\right)^{\frac{2}{\alpha}} \int^\infty_0 t^{-\frac{2}{\alpha}} \exp\left(-t\right) {\rm{d}} t \int^\infty_0 \frac{1}{\alpha} \theta^{\frac{2}{\alpha} - 1} \exp\left(-\theta\right) {\rm{d}} \theta \right)\\
= & \exp\left( -\pi \lambda_{\cal{D}} {\mathbb{P}}^{{\cal{D}}_1} R^2 \left(\xi^{\rm{D}}\right)^{\frac{2}{\alpha}} \int^\infty_0 t^{-\frac{2}{\alpha}} \exp\left(-t\right) {\rm{d}} t \int^\infty_0 \theta^{\frac{2}{\alpha}} \exp\left(-\theta\right) {\rm{d}} \theta \right) \\
\mathop = \limits^{(c)} & \exp\left( - \frac{\pi \lambda_{\cal{D}} {\mathbb{P}}^{{\cal{D}}_1} R^2 \left(\xi^{\rm{D}}\right)^{\frac{2}{\alpha}}}{{\rm{sinc}}\left(\frac{2}{\alpha}\right)} \right),
\end{aligned}
\end{equation}
where $(a)$ follows from the Rayleigh fading of $g_{d'd}$, $(b)$ follows the probability generating functionals (PGFL) of PPP~\cite{9516701}, and $(c)$ is obtained according to the reflection formula in Pi function (i.e., $\Pi\left(z\right)\Pi\left(-z\right) = \frac{\pi z}{\sin\left(\pi z\right)} = \frac{1}{{\rm{sinc}}\left(z\right)}$ with $\Pi\left(z\right) = z\Gamma\left(z\right) = \int_0^\infty t^z \exp\left(-t\right) {\rm{d}} t$ being the Pi function and $\Gamma\left(z\right)$ being the gamma function\footnote{Please refer to~\cite{havil2003gamma} and (43) in~\cite{7056528} for the definition and details.}). Similarly, the third term of the final row of~(\ref{eq:Prob_SINR_specific}) can be derived as follows:
\begin{equation}\label{eq:Prob_SINR_specific_third_term}
\begin{aligned}
&{\mathbb{E}}_{S_d^{{\cal{J}}}} \left[\exp \left( - \frac{R^{\alpha} \xi^{\rm{D}}} {p^{\rm{D}}} S_d^{{\cal{J}}}\right) \right] = {\mathbb{E}} \left[\exp \left( - \frac{R^{\alpha} \xi^{\rm{D}}} {p^{\rm{D}}} p^{\rm{J}} \sum_{j \in \Phi_{\left\{\left.{\cal{J}} \right| d \right\}}} g_{j d} r_{j d}^{-\alpha} \right) \right]\\
\mathop = \limits^{(\ref{eq:Prob_SINR_specific_second_term})}& \exp\left( - \frac{\pi \lambda_{\cal{J}} R^2 \left(\frac{p^{\rm{J}}}{p^{\rm{D}}}\xi^{\rm{D}}\right)^{\frac{2}{\alpha}}}{{\rm{sinc}}\left(\frac{2}{\alpha}\right)} \right).
\end{aligned}
\end{equation}

\section{Secrecy Outage Probability for D2D Network}
\label{app:Secrecy_outage_d2d}

According to~(\ref{eq:Prob_SecrecyOutage}) and conditioned on ${\cal{D}}_1$, the secrecy outage probability for D2D transmitter $d$ is  
\begin{equation}\label{eq:Prob_Secrecy_outage_specific}
\begin{aligned}
& {\mathbb{P}} \left[\left.{\rm{SINR}}_{d \left.w\right|_d} \left(p^{\rm{D}}, p^{\rm{J}}\right) \ge \xi^{\rm{W}} \right|{\cal{D}}_1\right] = {\mathbb{P}} \left[ \frac{p^{\rm{D}} g_{d \left.w\right|_d} r_{d \left.w\right|_d}^{-\alpha}}{ S_{\left.w\right|_d}^{{\cal{D}}_1} + S_{\left.w\right|_d}^{{\cal{J}}} + N_{\left.w\right|_d}}\ge \xi^{\rm{W}} \right]\\
= & {\mathbb{P}} \left[g_{d \left.w\right|_d} \ge \frac{\xi^{\rm{W}} r_{d \left.w\right|_d}^{\alpha}} {p^{\rm{D}}} \left(S_{\left.w\right|_d}^{{\cal{D}}_1} + S_{\left.w\right|_d}^{{\cal{J}}} + N_{\left.w\right|_d}\right) \right]\\ 
= & {\mathbb{E}} \left[\exp \left( - \frac{\xi^{\rm{W}} r_{d \left.w\right|_d}^{\alpha}} {p^{\rm{D}}} \left(S_{\left.w\right|_d}^{{\cal{D}}_1} + S_{\left.w\right|_d}^{{\cal{J}}} + N_{\left.w\right|_d}\right)\right) \right]\\ 
= & {\mathbb{E}}_{r_{d \left.w\right|_d}} \left[\exp \left( - \frac{\xi^{\rm{W}} r_{d \left.w\right|_d}^{\alpha}} {p^{\rm{D}}} N_{\left.w\right|_d}\right) {\mathbb{E}}_{S_{\left.w\right|_d}^{{\cal{D}}_1}} \left[\exp \left( - \frac{\xi^{\rm{W}} r_{d \left.w\right|_d}^{\alpha}} {p^{\rm{D}}} S_{\left.w\right|_d}^{{\cal{D}}_1}\right) \right]\right. \\
& \left. \times {\mathbb{E}}_{S_{\left.w\right|_d}^{{\cal{J}}}} \left[\exp \left( - \frac{\xi^{\rm{W}} r_{d \left.w\right|_d}^{\alpha}} {p^{\rm{D}}} S_{\left.w\right|_d}^{{\cal{J}}}\right) \right]\right],
\end{aligned}
\end{equation}
where the probability density function (PDF) of $r_{d\left.w\right|_d}$ is
\begin{equation}\label{eq:pdf_r_d_wd}
f_{r_{d\left.w\right|_d}}\left(r\right) = \exp\left(-\lambda_{\cal{W}}\pi r^2\right) 2\lambda_{\cal{W}}\pi r,
\end{equation}
\begin{equation}
{\mathbb{E}}_{S_{\left.w\right|_d}^{{\cal{D}}_1}} \left[\exp \left( - \frac{\xi^{\rm{W}} r_{d \left.w\right|_d}^{\alpha}} {p^{\rm{D}}} S_{\left.w\right|_d}^{{\cal{D}}_1}\right) \right] \mathop = \limits^{(\ref{eq:Prob_SINR_specific_second_term})} \exp\left( - \frac{\pi \lambda_{\cal{D}} {\mathbb{P}}^{{\cal{D}}_1} r_{d \left.w\right|_d}^2 \left(\xi^{\rm{W}}\right)^{\frac{2}{\alpha}}}{{\rm{sinc}}\left(\frac{2}{\alpha}\right)} \right),
\end{equation}
and 
\begin{equation}
{\mathbb{E}}_{S_{\left.w\right|_d}^{{\cal{J}}}} \left[\exp \left( - \frac{\xi^{\rm{W}} r_{d \left.w\right|_d}^{\alpha}} {p^{\rm{D}}} S_{\left.w\right|_d}^{{\cal{J}}}\right) \right] \mathop = \limits^{(\ref{eq:Prob_SINR_specific_second_term})} \exp\left( - \frac{\pi \lambda_{\cal{J}} r_{d \left.w\right|_d}^2 \left(\frac{p^{\rm{J}}}{p^{\rm{D}}}\xi^{\rm{W}}\right)^{\frac{2}{\alpha}}}{{\rm{sinc}}\left(\frac{2}{\alpha}\right)} \right).
\end{equation}

\section{False Alarm Probability for Warden}
\label{app:ProbFA}

According to~(\ref{eq:ProbFA}) and conditioned on ${\cal{D}}_0$, the FA probability for warden $\left.w\right|_d$ is 
\begin{equation}\label{eq:ProbFA_expression}
\begin{aligned}
& {\mathbb{P}}^{\text{FA}}_{\left.w\right|_d} \left(p^{\rm{D}}, p^{\rm{J}}, \tau\right) = {\mathbb{P}} \left[\left. y_{\left.w\right|_d} > \tau \right| {\cal{D}}_0 \right]\\
= & {\mathbb{P}} \left[S_{\left.w\right|_d}^{{\cal{D}}_1} + S_{\left.w\right|_d}^{{\cal{J}}} > \tau - N_{\left.w\right|_d} \right] = 1 - F_{S_{\left.w\right|_d}^{{\cal{D}}_1} + S_{\left.w\right|_d}^{{\cal{J}}}} \left(\tau - N_{\left.w\right|_d}\right).
\end{aligned}
\end{equation}
The cumulative distribution function (CDF) of $S_{\left.w\right|_d}^{{\cal{D}}_1} + S_{\left.w\right|_d}^{{\cal{J}}}$ can be obtained as follows:
\begin{enumerate}
\item We first calculate the Laplace transform of $S_{\left.w\right|_d}^{{\cal{D}}_1} + S_{\left.w\right|_d}^{{\cal{J}}}$ as 
\begin{equation}\label{eq:laplace_transform_SD1_SJ}
\begin{aligned}
& {\mathcal{L}} \left\{S_{\left.w\right|_d}^{{\cal{D}}_1} + S_{\left.w\right|_d}^{{\cal{J}}}\right\}\left(s\right) = {\mathbb{E}}\left[\exp\left(-s\left(S_{\left.w\right|_d}^{{\cal{D}}_1} + S_{\left.w\right|_d}^{{\cal{J}}}\right)\right)\right] \\
= & {\mathbb{E}}\left[\exp\left(-s S_{\left.w\right|_d}^{{\cal{D}}_1}\right)\right] {\mathbb{E}}\left[\exp\left(-s S_{\left.w\right|_d}^{{\cal{J}}}\right)\right] \\
\mathop = \limits^{(\ref{eq:Prob_SINR_specific_second_term})} & \exp\left( - \frac{\pi \lambda_{\cal{D}} {\mathbb{P}}^{{\cal{D}}_1} \left(s p^{\rm{D}}\right)^{\frac{2}{\alpha}}}{{\rm{sinc}}\left(\frac{2}{\alpha}\right)} \right) \exp\left( - \frac{\pi \lambda_{\cal{J}} \left(s p^{\rm{J}}\right)^{\frac{2}{\alpha}}}{{\rm{sinc}}\left(\frac{2}{\alpha}\right)} \right).
\end{aligned}
\end{equation}

\item The CDF of $S_{\left.w\right|_d}^{{\cal{D}}_1} + S_{\left.w\right|_d}^{{\cal{J}}}$ is the inverse Laplace transform of ${\mathcal{L}} \left\{S_{\left.w\right|_d}^{{\cal{D}}_1} + S_{\left.w\right|_d}^{{\cal{J}}}\right\}\left(s\right)$ and can be obtained by using the Bromwich inversion theorem in Chapter~2 of~\cite{cohen2007numerical} as follows:
\begin{equation}\label{eq:CDF_SD1_SJ}
\begin{aligned}
F_{S_{\left.w\right|_d}^{{\cal{D}}_1} + S_{\left.w\right|_d}^{{\cal{J}}}} \left(t\right) = & {\mathcal{L}}^{-1}\left\{{\mathcal{L}} \left\{S_{\left.w\right|_d}^{{\cal{D}}_1} + S_{\left.w\right|_d}^{{\cal{J}}}\right\}\left(s\right)\right\} \left(t\right)\\
= & 1 - \frac{1}{\pi} \int_0^\infty \frac{1}{\theta} \sin\left(\nu \theta^{\frac{2}{\alpha}} \sin\left(\frac{2\pi}{\alpha}\right) \right) \exp\left(- \nu \theta^{\frac{2}{\alpha}} \cos\left(\frac{2\pi}{\alpha}\right) - t \theta\right) {\rm{d}} \theta,
\end{aligned}
\end{equation}
where $\nu = \frac{\pi}{{\rm{sinc}}\left(\frac{2}{\alpha}\right)} \left(\lambda_{\cal{D}} {\mathbb{P}}^{{\cal{D}}_1} \left(p^{\rm{D}}\right)^{\frac{2}{\alpha}} + \lambda_{\cal{J}} \left(p^{\rm{J}}\right)^{\frac{2}{\alpha}}\right)$.
\end{enumerate}

\section{Miss Detection Probability for Warden}
\label{app:ProbMD}

According to~(\ref{eq:ProbMD}) and conditioned on ${\cal{D}}_1$, the MD probability for warden $\left.w\right|_d$ is 
\begin{equation}\label{eq:ProbMD_expression}
\begin{aligned}
{\mathbb{P}}^{\rm{MD}}_{\left.w\right|_d} \left(p^{\rm{D}}, p^{\rm{J}}, \tau\right) = & {\mathbb{P}} \left[\left. y_{\left.w\right|_d} < \tau \right| {\cal{D}}_1 \right] = {\mathbb{P}} \left[ p^{\rm{D}} g_{d\left.w\right|_d} r_{d\left.w\right|_d}^{-\alpha} + S_{\left.w\right|_d}^{{\cal{D}}_1} + S_{\left.w\right|_d}^{{\cal{J}}} + N_{\left.w\right|_d} < \tau \right]\\
= & \int_0^{\tau - N_{\left.w\right|_d}} f_{p^{\rm{D}} g_{d\left.w\right|_d} r_{d\left.w\right|_d}^{-\alpha}} \left(t\right) F_{S_{\left.w\right|_d}^{{\cal{D}}_1} + S_{\left.w\right|_d}^{{\cal{J}}}} \left(\tau - N_{\left.w\right|_d} - t\right) {\rm{d}} t,
\end{aligned}
\end{equation}
where $F_{S_{\left.w\right|_d}^{{\cal{D}}_1} + S_{\left.w\right|_d}^{{\cal{J}}}} \left(\cdot\right)$ is given in~(\ref{eq:CDF_SD1_SJ}) and 
\begin{equation}
\begin{aligned}
& f_{p^{\rm{D}} g_{d \left.w\right|_d} r_{d \left.w\right|_d}^{-\alpha}} \left(t\right) = \frac{{\rm{d}}}{{\rm{d}} t}F_{ p^{\rm{D}} h_{d \left.w\right|_d} r_{d \left.w\right|_d}^{-\alpha}} \left(t\right) = \frac{{\rm{d}}}{{\rm{d}} t}{\mathbb{P}} \left[p^{\rm{D}} h_{d \left.w\right|_d} r_{d \left.w\right|_d}^{-\alpha} \le t\right]\\
= & \frac{{\rm{d}}}{{\rm{d}} t}{\mathbb{P}} \left[ h_{d \left.w\right|_d} \le \frac{tr_{d \left.w\right|_d}^{\alpha}}{p^{\rm{D}}} \right] = \frac{{\rm{d}}}{{\rm{d}} t}{\mathbb{E}}_{r_{d \left.w\right|_d}} \left[1 -  \exp\left( -\frac{tr_{d \left.w\right|_d}^{\alpha}}{p^{\rm{D}}} \right) \right] \\
= & \int_0^\infty f_{r_{d\left.w\right|_d}}\left(r\right) \exp\left( -\frac{t r^{\alpha}}{p^{\rm{D}}} \right) \frac{r^{\alpha}}{p^{\rm{D}}} {\rm{d}} r
\end{aligned}
\end{equation}
with $f_{r_{d\left.w\right|_d}}\left(r\right)$ given in~(\ref{eq:pdf_r_d_wd}).


\end{appendices}

\bibliography{bibfile}

\end{document}